\documentclass[a4paper,11pt]{article}

\usepackage{physics}
\usepackage{color}
\usepackage[usenames,dvipsnames]{xcolor}
\usepackage{graphicx}
\usepackage{amsmath}
\usepackage{caption}
\usepackage{siunitx}
\usepackage{booktabs}
\usepackage{tabularx}
\usepackage{multirow}
\usepackage{multicol}
\usepackage{transparent}
\usepackage{placeins}

\usepackage{relsize}

\pdfoutput=1 % if your are submitting a pdflatex (i.e. if you have
\usepackage{jheppub} % for details on the use of the package, please
\hypersetup{%
  colorlinks=true, linktocpage=true, pdfstartpage=1, pdfstartview=FitV,
  breaklinks=true, pageanchor=true,
  pdfpagemode=UseNone,
  plainpages=false, bookmarksnumbered, bookmarksopen=true, bookmarksopenlevel=1,
  hypertexnames=true, pdfhighlight=/O,
  urlcolor=RoyalBlue, linkcolor=ForestGreen, citecolor=RoyalBlue,
  pdftitle={Renormalisation group analysis of scalar Leptoquark
    couplings addressing flavour anomalies},
  pdfauthor={Marco Fedele, Ulrich Nierste, Felix Wuest},
  pdfsubject={},
  pdfkeywords={},
  pdfcreator={pdfLaTeX},
  pdfproducer={LaTeX with hyperref}
}

\newcommand{\lt}{\left}
\newcommand{\rt}{\right}
\newcommand{\gev}{\,\mbox{GeV}}

\newcommand{\fig}[1]{Fig.~\ref{#1}}

\newcommand{\eq}[1]{Eq.~(\ref{#1})}

\preprint{TTP23-027, P3H-23-048}

\author{Marco Fedele,}
\author{Ulrich Nierste,}
\author{Felix Wuest}

\affiliation{Institute for Theoretical Particle Physics, Karlsruhe Institute of Technology (KIT),\\
  Wolfgang-Gaede-Str. 1, 76131 Karlsruhe, Germany}

\emailAdd{marco.fedele@kit.edu}
\emailAdd{ulrich.nierste@kit.edu}
\emailAdd{felix.wuest@student.kit.edu}

\abstract{Leptoquarks with masses between 2 TeV and 50 TeV are commonly
  invoked to explain deviations between data and Standard-Model (SM)
  predictions of several observables in the decays $b\to c\tau \bar\nu$
  and $b\to s \ell^+\ell^-$ with $\ell=e,\mu$. While Leptoquarks appear
  in theories unifying quarks and leptons, the corresponding
  unification scale $M_{QLU}$ is typically many orders of magnitude
  above this mass range. We study the case that the mass gap between the
  electroweak scale and $M_{QLU}$ is only populated by scalar
  Leptoquarks and SM particles, restricting ourselves to scenarios
  addressing the mentioned flavour anomalies, and determine the
  renormalisation-group evolution of Leptoquark couplings to fermions
  below $M_{QLU}$. In the most general case, we consider three SU(2) triplet Leptoquarks $S_3^\ell$,
  $\ell=e,\mu,\tau$, which couple quark doublets to the lepton doublet
  $(\nu_\ell,\ell^-)$ to address the $b\to s \ell^+\ell^-$ anomalies.
  In this case, we find a scenario in which the
  Leptoquark couplings to electrons and muons are driven to the same
  infrared fixed point, so that lepton flavour universality emerges
  dynamically. However, the corresponding fixed point for the couplings 
  to taus is necessarily opposite in sign, leading to a unique signature
  in $b \to s\tau^+\tau^-$.
  For $b\to c\tau \bar\nu$ we complement these with either an SU(2)
  singlet $S_1^\tau$ or doublet $R_2^\tau$ and study further the cases
  that also these Leptoquarks come in three replicas.  The fixed point
  solutions for the $S_3^\ell$ couplings explain the
  $b\to s \ell^+\ell^-$ data for $S_3^{e,\mu}$ masses between
  14 and 15 TeV, according to the scenario.  
  $b\to c\tau \bar\nu$ data can only be fully
  explained by couplings exceeding their fixed-point values and evolving
  into Landau poles at high energies, so that one can place an upper
  bound on $M_{QLU}$ between $10^{8}$ and $10^{11}$ GeV.  }

\title{Renormalisation group analysis of scalar Leptoquark couplings addressing
    flavour anomalies: emergence of lepton-flavour universality}

\begin{document}

\maketitle
\flushbottom

\section{Introduction}\label{sec:intro}
Several measured branching ratios driven by the quark decay
$b\to s \mu^+\mu^-$ show a deficit of events in the kinematic region
with $q^2\leq 8\gev$, where $q^2$ is the invariant mass of the lepton
pair~\cite{LHCb:2014cxe,LHCb:2015wdu,LHCb:2021zwz}, if confronted with the Standard-Model (SM) prediction of
Refs.~\cite{Khodjamirian:2010vf,Khodjamirian:2012rm}. Also the
observable $P_5^\prime$ parametrising an angular distribution in
$B\to K^* \mu^+\mu^-$ follows this pattern~\cite{LHCb:2013ghj,LHCb:2015svh,LHCb:2020lmf,LHCb:2020gog}.\footnote{Refs.~\cite{Khodjamirian:2010vf,Khodjamirian:2012rm} employ  QCD sum rules, a method in which the contribution from excited hadrons to correlation functions is calculated perturbatively (``quark-hadron duality''). The uncertainty associated with this step is hard to quantify and critics suggested this as the source of the discrepancy.} 
In a 2022 reanalysis of LHCb data for the lepton flavour universality violating (LFUV) ratios~\cite{hep-ph/0310219}
\begin{eqnarray}
  R_{K^{(*)}}
  & \equiv &
             \frac{B(B\to K^{(*)} \mu^+\mu^-)}{B(B\to K^{(*)} e^+e^-)}
\end{eqnarray}
has resulted in values compatible with the SM predictions
$ R_{K^{(*)}}\simeq 1$~\cite{LHCb:2022qnv,LHCb:2022zom}. Thus, while the previous results of $ R_{K^{(*)}} \sim 0.8$ were hinting at the possibility of LFUV up to the 20\% level, the current situation has strongly changed and violation of LFU is no longer preferred by data, although technically still allowed with reduced size. Therefore, if beyond-SM (BSM)
physics is invoked to explain the $b\to s \mu^+\mu^-$, it will couple
with similar strengths to muons and electrons. 

Another long-standing flavour anomaly is related to $b \to c\tau \nu$
decays and observed in the ratios
\begin{eqnarray}
  R_{D^{*}}
  & \equiv &
             \frac{B(B\to D^{(*)} \tau \nu )}{B(B\to D^{(*)} \ell\nu)},
             \qquad\qquad \ell=e,\mu .
\end{eqnarray}
While BaBar and Belle have measured both ratios jointly, early LHCb
measurements could only determine $R_{D^{*}}$. While all measurements
have always been very consistent concerning $R_{D^{*}}$, there is some
tension between the large 2012 BaBar value for $R_D$~\cite{BaBar:2012obs} and the
corresponding 2019 Belle measurement with a smaller, SM-like
result to the level expected by statistical fluctuation~\cite{Belle:2019rba}.  In 2022 LHCb has presented a combined $R_{D}-R_{D^{*}}$ measurement which has increased the overall consistency  
among all experimental results~\cite{LHCb:2023zxo}. HFLAV combines six
measurements \cite{BaBar:2012obs,Belle:2019rba,LHCb:2023zxo,Belle:2015qfa,Belle:2016ure,LHCbSem} to~\cite{HFLAV:2022pwe}
\begin{eqnarray}
   R_{D}^{\rm exp} &=&  0.358 \pm 0.025 \pm 0.012\,, \qquad\qquad 
             R_{D^{*}}^{\rm exp} \;=\;  0.285 \pm 0.010 \pm 0.008\,, \label{rdexp}
\end{eqnarray} 
which have to be compared with the SM predictions of~\cite{Bigi:2016mdz,Bernlochner:2017jka,Jaiswal:2017rve,Gambino:2019sif,Bordone:2019vic,Martinelli:2021onb}
\begin{eqnarray}
   R_{D} &=& 0.298 \pm 0.004\,,\qquad\qquad
   R_{D^{*}} \;=\;  0.254 \pm 0.005\,,
\end{eqnarray}
entailing a discrepancy with \eq{rdexp} of 3.2$\sigma$.
Better
measurements of $D^*$ and $\tau$ polarisations can discriminate between
different BSM explanations of $R_{D^{(*)}}$ \cite{Blanke:2018yud,Blanke:2019qrx}. 
The ratio  $R_{\Lambda_c}\equiv B(\Lambda_b\to\Lambda_c \tau \nu )/
B(\Lambda_b\to\Lambda_c\ell\nu)$ contains redundant information to 
$R_{D^{*}}$ in any model of New Physics (NP)
\cite{Blanke:2018yud,Blanke:2019qrx} and must move upward in
future measurements from its 2022 value $R_{\Lambda_c}^{\rm LHCb}=
0.242\pm 0.026\pm 0.040 \pm 0.059$~\cite{LHCb:2022piu} to 
$R_{\Lambda_c} = 0.39\pm 0.05$~\cite{Fedele:2022iib} if $R_{D^{(*)}}^{\rm exp} $ in \eq{rdexp} are
correct.
\begin{figure}
  \includegraphics[width=\textwidth]{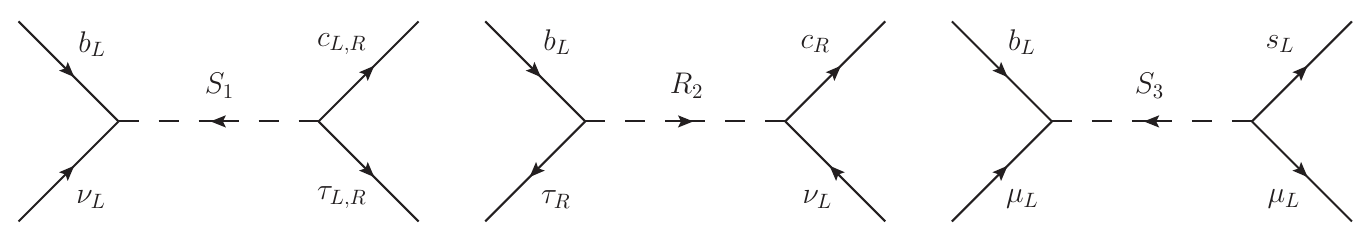}
  \caption{Contributions of scalar Leptoquarks to anomalous quark decays.\label{fig:bdec}
  }
\end{figure}

Leptoquarks (LQs) are the most popular particle species postulated to remedy
the flavour anomalies. While giving an exhaustive list of references on this topic goes beyond the scope of this paper, a selection of the most prominent and innovative models involving LQs and addressing $b\to s$ and $b\to c$ anomalies can be found in Refs~\cite{Sakaki:2013bfa,Hiller:2014yaa,Dorsner:2016wpm,
  Dumont:2016xpj, Li:2016vvp,Hiller:2016kry,Bhattacharya:2016mcc,Chen:2017hir,  Crivellin:2017zlb,Jung:2018lfu,Aydemir:2019ynb,Popov:2019tyc,Crivellin:2019dwb,
  Iguro:2020keo}, and references therein.
In this paper we focus on scalar LQs, which can be consistently
added to the SM particle content. That is, their mass $M_{\rm LQ}$ is
much below the scale $M_{\rm QLU}$ determining the masses of the
remaining particles of some complete theory of quark-lepton unification (QLU)
and the effects of the latter particles decouple for
$M_{\rm QLU}\to\infty$. By contrast, a vector LQ  with mass
$M_{\rm LQ}\ll M_{\rm QLU}$ corresponds to a non-decoupling scenario
unless the Higgs sector responsible for its mass is taken into account as
well. Flavour anomalies are addressed with the scalar LQs $S_1$, $R_2$
and $S_3$, denoting SU(2) singlet, doublet and triplet respectively, see \fig{fig:bdec} for sample diagrams. The
combinations $(S_1,S_3)$ or $(R_2,S_3)$ can simultaneously cure $b\to
c\tau \bar\nu$ and $b\to s \ell^+\ell^-$, with the caveat that one needs
more than one copy of some SU(2) representations as explained below in
Sec.~\ref{sec:Leptoquarks}. The former scenario also affects the decay
$b\to s \nu \bar\nu$ which is currently probed at the Belle II experiment. 
In both scenarios one can find large effects on the anomalous magnetic
moment of the muon \cite{Athron:2021iuf}.

The presence of a sizable mass gap between $M_{\rm LQ}$ and
$M_{\rm QLU}$ opens the possibility to study the renormalisation group
(RG) to find generic predictions for the low-energy parameters without
specifying details of the complete theory valid at $M_{\rm QLU}$.  The
prototypical example for such a study is gauge coupling unification,
which can be assessed from the SM beta functions alone, without knowing
the parameters of the grand unified theory valid at the high
scale. Indeed, the ``near miss'' of these running couplings nurtures the hope to
find new particles in the reach of current particle colliders, because
they change the slope of the beta functions.  Another
opportunity of RG analyses is the possibility to find infrared (IR) (quasi-)
fixed points (FP) of parameters. Such studies have been pioneered in
Ref.~\cite{Pendleton:1980as} for the top Yukawa coupling, aiming at a
prediction of the top mass. In this paper we derive and study the RG
equations for LQ Yukawa couplings and  SM gauge couplings.%\footnote{For a study of the implications of asymptotically safe quantum gravity on ultraviolet (UV) FP, see Ref.~\cite{Kowalska:2020gie}.}

The paper is organised as follows: In Sec.~\ref{sec:efts} we report the effective
Hamiltonians employed to describe $B$ Meson decays in and beyond the SM,
and summarize the current status of bounds on the NP couplings 
from the latest global fits. Sec.~\ref{sec:Leptoquarks}
reviews some basics and assesses the implications of low-energy data
on the flavour pattern of the LQ Lagrangian. In Sec.~\ref{sec:rge}
we present the RG equations (RGE) of the LQ couplings first
in a fully general theory and then specifically for the
scenarios which can explain the flavour anomalies. Sec.~\ref{sec:fixedpoints}
discusses the RGE FPs and their implications. Finally we
conclude in Sec.~\ref{sec:summary}.

%%%%%%%%%%%%%%%%%%%%%%%%%%%%%%%%%

\section{Effective Hamiltonians for \texorpdfstring{$B$}{B} Meson Decays}\label{sec:efts}
It is customary to describe the decays of $B$ mesons in the SM by means of effective field theories (EFTs), obtained after integrating out the top quark, the heavy gauge bosons $Z$ and $W$, and the Higgs field. This approach is particularly helpful in the presence of BSM physics as well. Indeed, the low-scale footprints of any heavy degree of freedom can be parametrized at the $B$ meson decay scale as shifts to the Wilson coefficients (WCs), describing the short-distance effects associated to all the fields integrated out of the theory. Therefore, after performing fits to all the available experimental data, it is possible to obtain bounds on the NP effects in a model independent way. These bounds can be then translated into constraints on any given model once the matching between the EFT and the desired BSM theory is performed. We will give the results of such matching for the relevant LQs in Sec.~\ref{sec:Leptoquarks}.

The effective Hamiltonian employed to describe $b\to s\ell^+\ell^-$ transitions reads
%%%%%%%%%%%%
\begin{equation}\label{eq:Heff_bsll}
\mathcal{H}_{\rm eff}^{\ell\ell} \supset \frac{4 G_F}{\sqrt{2}}V_{tb}V_{ts}^\ast \Big{(}C_9^\ell\mathcal{O}_9^\ell+C_{10}^\ell\mathcal{O}_{10}^\ell\Big{)} + \text{h.c.}\,,
\end{equation}
%%%%%%%%%%%%
where we focus on the phenomenologically relevant operators
%%%%%%%%%%%%
\begin{equation}
\mathcal{O}_9^\ell = \frac{\alpha_{\rm em}}{4\pi}(\bar{s}\gamma_\mu P_L b)(\bar{\ell}\gamma^\mu \ell)\,, \qquad\qquad  
\mathcal{O}_{10}^\ell =\frac{\alpha_{\rm em}}{4\pi}(\bar{s}\gamma_\mu P_L b)(\bar{\ell}\gamma^\mu \gamma_5 \ell)\,.
\end{equation}
%%%%%%%%%%%%
Here $G_F$ is the Fermi constant, $V_{tb}$ and $V_{ts}$ are elements of the Cabibbo-Kobayashi-Maskawa (CKM) matrix, $\alpha_{\rm em}$ is the fine structure constant and $P_{L,R} = (1 \mp \gamma_5)/2$. In the SM, the WCs are LFU and at the renormalization scale $\mu\equiv\mu_b=4.8$ GeV equal to $C_9^\ell(\mu_b)\simeq4.1$ and $C_{10}^\ell(\mu_b)\simeq-4.3$, respectively. It is also useful to define the quantity  $C_{L}^{\ell}\equiv C_9^\ell=-C_{10}^\ell$.

As anticipated in the Introduction, the latest experimental results concerning $R_{K^{(*)}}$~\cite{LHCb:2022qnv,LHCb:2022zom} require NP effects to be LFU, if one wants to address the discrepancies in $b\to s\mu^+\mu^-$ transitions by means of BSM physics. Defining therefore these additional contributions as ${C_9^{\rm U}\equiv C_9^e=C_9^\mu}$ and ${C_{L}^{\rm U}\equiv C_L^e =C_L^\mu}$, the most likely results found by the latest global fits~\cite{Ciuchini:2022wbq,Greljo:2022jac,Alguero:2023jeh} are 
\begin{equation}\label{eq:bsll_gf}
\begin{aligned}
    \textbf{I)}&\qquad C_9^{\rm U}(\mu_b) \sim -1\,,\\
    \textbf{II)}&\qquad C_L^{\rm U}(\mu_b) \sim -0.4\,.
\end{aligned}
\end{equation}
As we will see in the next Section, the WCs configuration found in scenario $\textbf{II)}$ arises in the presence of $S_3$ LQs coupling equally to electron and muons.

It is interesting to notice that $b\to s\nu\bar\nu$ transitions can be described by an effective Hamiltonian closely related to the one given at Eq.~\eqref{eq:Heff_bsll}, namely
%%%%%%%%%%%%
\begin{equation}\label{eq:Heff_bsnunu}
\mathcal{H}_{\rm eff}^{\nu\bar\nu} \supset - \frac{4 G_F}{\sqrt{2}}V_{tb}V_{ts}^\ast C_{\nu\bar\nu}^\ell\mathcal{O}_{\nu\bar\nu}^\ell + \text{h.c.}\,,
\end{equation}
%%%%%%%%%%%%
where we have introduced the neutrino operator
%%%%%%%%%%%%
\begin{equation}
\mathcal{O}_{\nu\bar\nu}^\ell = \frac{\alpha_{\rm em}}{4\pi}(\bar{s}\gamma_\mu P_L b)(\bar{\nu}_\ell\gamma^\mu(1-\gamma_5) \nu_\ell)\,.
\end{equation}
%%%%%%%%%%%%
Since experiment cannot distinguish neutrino flavours, the sum over all flavours appears in the ratio of the branching fraction and its SM prediction~\cite{Buras:2014fpa}:
%%%%%%%%%%%%
\begin{equation}\label{eq:RKnunu}
R_{K^{(*)}}^{\nu\bar\nu} = \frac{{\mathcal B}^{\rm exp}(B\to K^{(*)} \nu\bar\nu)}{{\mathcal B}^{\rm SM}(B\to K^{(*)} \nu\bar\nu)} =  
\frac{(C_{\nu\bar\nu}^{\rm SM} + C_{\nu\bar\nu}^e)^2+(C_{\nu\bar\nu}^{\rm SM} + C_{\nu\bar\nu}^\mu)^2+(C_{\nu\bar\nu}^{\rm SM} + C_{\nu\bar\nu}^\tau)^2}{3(C_{\nu\bar\nu}^{\rm SM})^2}  \,,
\end{equation}
%%%%%%%%%%%%
where $C_{\nu\bar\nu}^{\rm SM}(\mu_b)\simeq-6.35$. On the one hand the current experimental limits in the $K^*$ channel are set by the Belle collaboration~\cite{Grygier:2017tzo}, and read at $90 \%$ C.L. $R_{K^*}^{\nu \bar \nu} < 2.7$. On the other hand, concerning the $K$ channel, the Belle II collaboration recently reported the first observation of this decay, which was found to be $2.8\sigma$ above its SM prediction~\cite{BelleIISemKnunu}, and corresponding to $R_{K}^{\nu \bar \nu} = 2.8 \pm 0.8$ when combined with previous measurements. In the case where NP couples to only one lepton flavour, these bounds imply at the $2\sigma$ level
%%%%%%%%%%%%
\begin{equation}\label{eq:bsnunu_gf}
-9 \lesssim C_{\nu\bar\nu}^{\rm NP} \lesssim -1.7 \quad\cup\quad 14 \lesssim C_{\nu\bar\nu}^{\rm NP} \lesssim 22 \,,
\end{equation}
%%%%%%%%%%%%
where $C_{\nu\bar\nu}^{\rm NP}$ represents any of $C_{\nu\bar\nu}^{e,\mu,\tau}$.
An upcoming measurement by the Belle II collaboration in the $B\to K^{*} \nu\bar\nu$ channel is expected as well, which will provide further constraints on $C_{\nu\bar\nu}^{\rm NP}$~\cite{Belle-II:2018jsg}.

The $b\to c \ell\nu$ transitions are described by the following effective Hamiltonian:
%%%%%%%%%%%%
\begin{equation}\label{eq:Heff_bclnu}
 {\cal H}_{\rm eff}^{\ell\nu} \supset \frac{4 G_F}{\sqrt{2}} V^{}_{cb} \big[(1+C_{V_L}^{\ell}) \mathcal{O}_{V_L}^{\ell} + C_{S_L}^{\ell} \mathcal{O}_{S_L}^{\ell} + C_{S_R}^{\ell} \mathcal{O}_{S_R}^{\ell}+C_{T}^{\ell} \mathcal{O}_{T}^{\ell}\big] + \text{h.c.}\,,
\end{equation}
%%%%%%%%%%%%
where we have introduced the operators
%%%%%%%%%%%%
\begin{equation}
\begin{aligned}
   \mathcal{O}_{V_L}^{\ell} &= \left(\bar c\gamma ^{\mu } P_L b\right)  \left(\bar {\ell} \gamma_{\mu } P_L \nu_{{\ell}}\right)\,, \qquad\qquad   
   \mathcal{O}_{S_L}^{\ell}  = \left( \bar c P_L b \right) \left( \bar {\ell} P_L \nu_{{\ell}}\right)\,,   \\
   \mathcal{O}_{S_R}^{\ell} &= \left( \bar c P_R b \right) \left( \bar {\ell} P_L \nu_{\ell}\right)\,, \qquad\qquad\qquad\;\,
   \mathcal{O}_{T}^{\ell}  = \left( \bar c \sigma^{\mu\nu}P_L  b \right) \left( \bar {\ell} \sigma_{\mu\nu} P_L \nu_{{\ell}}\right)\,,
\end{aligned}
\end{equation}
%%%%%%%%%%%%
with $\sigma_{\mu\nu}=\frac i2 [\gamma_\mu,\gamma_\nu]$. Given the normalization employed in Eq.~\eqref{eq:Heff_bclnu}, all the WCs there appearing are describing genuine NP effects. It is worth to mention that in our study we will not consider effects coming from the operator $\mathcal{O}_{V_R}$, which is obtained by replacing $P_L$ with $P_R$ in the quark bilinear of $\mathcal{O}_{V_L}$, as it is LFU at dimension-six in the SMEFT~\cite{Buchmuller:1985jz,Grzadkowski:2010es,Alonso:2014csa,Aebischer:2015fzz}. Moreover, we do not allow for effects coming from right-handed neutrinos.

The latest bounds on the NP WCs involved in $b \to c \ell\nu$ transitions, both in a model-independent way and for specific UV models, can be found, e.g., in Ref.~\cite{Iguro:2022yzr}. As detailed in the following Section, out of the several possible scenarios identified by the fit we focus here on the following scenarios, given at the renormalization scale $\mu_b$:
\begin{equation}\label{eq:bclnu_gf}
\begin{aligned}
    \textbf{A)}&\qquad C_{V_L}^\tau(\mu_b)\sim0.08\,,\\
    \textbf{B)}&\qquad C_{S_L}^\tau(\mu_b)=-8.9 C_T^\tau(\mu_b)\sim0.19\,,\\
    \textbf{C)}&\qquad C_{S_L}^\tau(\mu_b)=8.4 C_T^\tau(\mu_b)\sim \pm i0.58\,.
\end{aligned}
\end{equation}
Scenarios $\textbf{A)}$ and/or $\textbf{B)}$ can arise in the presence of a $S_1$ LQ coupled to taus, while $\textbf{C)}$ is instead a combination of WCs induced at the low scale by the presence of a $R_2$ LQ, coupling to taus.

%%%%%%%%%%%%%%%%%%%%%%%%%%%%%%%%%

\section{Theory of Leptoquarks}\label{sec:Leptoquarks}
The updated LHCb values for $R_{K^{(*)}}$ \cite{LHCb:2022qnv,LHCb:2022zom} imply that the NP interpretation of $b\to s\ell^+\ell^-$ data requires that 
both $b\to s \mu^+\mu^-$ and $b\to s e^+e^-$ receive NP
contributions with similar size~\cite{Ciuchini:2022wbq,Greljo:2022jac,Alguero:2023jeh}. As an immediate consequence, the $S_3$ LQ potentially
mediating these decays must come in two copies, $S_3^e$ and $S_3^\mu$,
each coupling only to the indicated lepton species. The reason why a single LQ cannot couple to both electrons and muons is the strong experimental bound on $\mu\to e$ conversion, which such a LQ would otherwise mediate. In the SM we observe an approximate $SU(2)^2$ flavour symmetry, corresponding to rotations of the charged right-handed fields $(l_{1R},l_{2R})$  and the left-handed 
doublets $(L_{1},L_{2})$ of the first two fermion generation. A priori 
the $S_3$ fields will couple to the weak eigenstates and the rotations 
of the latter into the flavour eigenstates $e_{L,R}$, $\mu_{L,R}$ 
(upon diagonalisation of the SM lepton Yukawa matrix) will lead to 
Leptoquarks coupling to both $e$ and $\mu$, which we must avoid. 
This rotation, however, is unphysical, if the LQ mass matrix is proportional to the unit matrix, in which case one 
finds $S_3^e$ and $S_3^\mu$  as desired. Mass-degenerate $S_3^e$ and $S_3^\mu$ mean that the LQ mass term in the Lagrangian also obeys an SU(2) flavour symmetry related to rotations of leptons in flavour space.\footnote{We remark that such flavor symmetry forbids the presence of Higgs-mediated LQ self-interaction terms such as $(S_3^{e,\dagger}S_3^\mu)(\Phi^\dagger\Phi)$, which we therefore do not allow in our Lagrangians. Besides, contributions to the Yukawa RGE from self-interaction terms arises only at the two-loop level, hence beyond the scope of this paper.}
Thus we conclude from the experimental evidence for $R_{K^{(*)}}\sim 1$ that Leptoquarks are part of the flavour puzzle and part (or even actors) of its explanation in term of approximate SU(2) symmetries. 

For the $b\to c\tau \nu$ anomalies one may employ $S_1$ or $R_2$ exchange, see \fig{fig:bdec}. For the former solution 
the $S_1$ coupling to $\bar c_L \tau^c_L$ comes with a coupling to $\bar s_L \nu_{\tau L}^c$ by SU(2) symmetry. 
This gives a large contribution to $b\to s \nu \bar\nu$, which could be mitigated by an $S_3^\tau$ contribution of opposite sign in an appropriate model~\cite{Crivellin:2017zlb}. Therefore the $(S_1,S_3)$ scenario could permit a significant enhancement of 
the branching ratio of $B\to K^{(*)} \nu \bar\nu $ currently studied at Belle II~\cite{Belle-II:2018jsg,BelleIISemKnunu}. The $R_2$ scenario can only successfully explain both $R(D)$ and $R(D^*)$ if the 
real part of the product of the $\bar\tau b_L$ and $\bar c\tau_R$ Yukawa couplings of $R_2$ is much smaller than the imaginary part (in the usual quark basis in which $V_{cb}$ is real) \cite{Becirevic:2018afm,Blanke:2018yud, Blanke:2019qrx,Iguro:2020keo}.

%%%%%%%%%%%%%%%%%%%%%%%%%%%%%%%%%

\subsection{Lagrangians}\label{sec:lag}
Let us here review the formalism employed to describe scalar LQs. In order to do so, we adopt for fermion fields $\psi$ the following formalism: $\psi_{L,R} = P_{L,R}\psi$, $\bar{\psi}=\psi^\dagger\gamma^0$ and $\psi^C=C\bar{\psi}^T$, where we have introduced $C=i\gamma^2\gamma^0$.

In the following we report the Lagrangians describing the interaction of scalar LQs with SM fields. We do not permit here diquark coupling of LQ, which would lead to dangerous and undesired proton decays~\cite{Dorsner:2016wpm}, and do not consider LQs coupling to right-handed neutrinos. Hence, we will focus only on five families of scalar LQs, each denoted by different quantum numbers relatively to the SM gauge group $(SU(3),SU(2),U(1))$~\cite{Buchmuller:1986zs}. In particular, we employ a fully general formalism, allowing in principle multiple copies for each LQ.

Before going into details for each LQ scenario we report here the generalization of the SM Yukawa Lagrangian to the case of $n_H$ scalar Higgs doublets $\Phi^a$, where $a=1,\dots,n_H$, with generic flavour structure. These theories are usually defined as generic $n_H$ Higgs doublet models (GNHDM), and can be described by the following Lagrangian:
%%%%%%%%%%
\begin{equation}
   \label{eq:yukawalagrangian}
   \mathcal{L}_{\Phi} = - Y^a_{u,\,ij} \Bar{Q}_{L,\,i}^l \epsilon^{lm} \Phi^{a,\,m}  u_{R,\,j} -  Y^a_{d,\,ij} \Bar{Q}_{L,\,i} \Phi^a d_{R,\,j}  - Y^a_{e,\,ij} \Bar{L}_{L,\,i} \Phi^a e_{R,\,j} + \text{h.c.}\,,
\end{equation}
%%%%%%%%%%
where $\epsilon^{lm}=(i\tau^2)^{lm}$, with $\tau^2$ being the second Pauli matrix. Moreover, $l,m=1,2$ are $SU(2)$ indices and $i,j=1,2,3$ are flavour indices. As stated above, we do not assume any particular flavour structure in the couplings among the several scalar Higgs doublets and the SM fields, namely each Higgs doublet $\Phi^a$ can couple with all SM fermions through the fully general coupling matrices $Y^a_{u,d,e}$.

Finally, we adopt the convention $g_1\equiv\sqrt{3/5}g'$, $g_2\equiv g$ and $g_3\equiv g_s$, with $g'$, $g$ and $g_s$ being the $U(1)$, $SU(2)$ and $SU(3)$ gauge couplings, respectively.

%%%%%%%%%%%%%%%%%%%%%%%%%%%%%%%%%

\subsubsection{Singlet Leptoquarks}
\label{sec:S1intro}

A scalar LQ $S_1 \equiv (\mathbf{\Bar{3}},\mathbf{1},1/3)$ interacts with the SM fields via the following Lagrangian:
%%%%%%%%%%%%
\begin{equation}
    \label{eq:S1yukawaL}
    \mathcal{L}_{\mathcal{Y}_{S_1}} \,  =  y_{1\,ij}^a \Bar{Q}^{C,\,l}_{L,\,i} S_1^a \epsilon^{lm} L_{L,\,j}^{m} + x_{1\,ij}^a \Bar{u}^{C}_{R,\,i} S_1^a e_{R,\,j} + \text{h.c.}\,.
\end{equation}
%%%%%%%%%%%%%
This Lagrangian describes all the coupling that are allowed for a weak singlet $S_1$, which can couple either to two left-handed SM fermions, or to two right-handed ones. Similarly to the convention adopted for the Higgs doublets, here and below the index $a$ is a family index employed to denote an arbitrary number of copies of a scalar LQ. This index can also be interpreted as a flavour index, analogously to the flavour indices $i,j$ of the SM fermion fields. The interaction between an $S_1^a$ LQ and the SM fields is mediated by arbitrary complex $3\times3$ Yukawa coupling matrices $y_{1}^a$ and $x_{1}^a$, connected to left-handed and right-handed fermions respectively.

On the other hand, the interaction among a scalar LQ $\tilde{S}_1 \equiv (\mathbf{\Bar{3}},\mathbf{1},4/3)$ and SM fields is described by
%%%%%%%%%%%%
\begin{equation}
    \label{eq:S1tyukawaL}
    \mathcal{L}_{\mathcal{Y}_{\tilde{S}_1}} \,  = \tilde x_{1\,ij}^a \Bar{d}^{C}_{R,\,i} \tilde{S}_1^a e_{R,\,j} + \text{h.c.}\,.
\end{equation}
%%%%%%%%%%%%%
Contrarily to $S_1$ in Eq.~\eqref{eq:S1yukawaL}, a weak singlet $\tilde S_1$ can only couple to two right-handed fields due to hypercharge conservation. This interaction is mediated by the arbitrary complex $3\times3$ Yukawa coupling matrix $\tilde x_{1}^a$.

The only scalar LQ which is going to be relevant for the phenomenological studies carried out in Sec.~\ref{sec:fixedpoints} is $S_1^\tau$, once non-vanishing values for the couplings $y_{1\,23}^\tau$, $y_{1\,33}^\tau$ and $x_{1\,23}^\tau$ are allowed. Indeed, it can contribute to $b\to c\tau\nu$ decays via~\cite{Angelescu:2018tyl}
%%%%%%%%%%%%
\begin{equation}\label{eq:S1_bc}
C_{S_L}^\tau(\mu_{\rm LQ}) = -4 C_T^\tau(\mu_{\rm LQ}) = - \frac{v^2}{4 V_{cb}}\frac{y_{1\,33}^\tau x_{1\,23}^{\tau\,*}}{M_{S_1^\tau}^2}\,, \qquad  
C_{V_L}^\tau(\mu_{\rm LQ}) = \frac{v^2}{4 V_{cb}}\frac{y_{1\,33}^\tau (V_{cs} y_{1\,23}^{\tau\,*} + V_{cb} y_{1\,33}^{\tau\,*})}{M_{S_1^\tau}^2}\,,
\end{equation}
%%%%%%%%%%%%%
at the matching scale $\mu_{\rm LQ}=M_{S_1^\tau}\sim 2$ TeV, with $v=246$ GeV. Notice that the relations among $C_{S_L}^\tau$ and $C_T^\tau$ is modified due to RGE effects once the coefficients are run down to the low scale, becoming $C_{S_L}^\tau(\mu_b) = -8.9 C_T^\tau(\mu_b)$~\cite{Gonzalez-Alonso:2017iyc,Aebischer:2018acj}. It is worth mentioning that, due to $SU(2)$ invariance, the presence of $y_{1\,33}^\tau$ and $y_{1\,23}^\tau$ implies a contribution to $b \to s\nu\bar\nu$ transitions as well, equal to~\cite{Buras:2014fpa}
%%%%%%%%%%%%
\begin{equation}
C_{\nu\bar\nu}^\tau = \frac{\pi v^2 }{V_{tb} V_{ts}^*\alpha_{\rm em}}\frac{y_{1\,33}^\tau y_{1\,23}^{\tau\,*}}{m_{S_1^\tau}^2}\,.
\end{equation}
%%%%%%%%%%%%%
Employing the results for scenarios $\textbf{B)}$ or $\textbf{A)}$ given in Eq.~\eqref{eq:bclnu_gf} at the decay scale (which therefore take into account the running effects from $\mu_{\rm LQ}=M_{S_1^\tau}$ to $\mu=\mu_b$) implies the following expected size for the NP parameters ratios, respectively: 
%%%%%%%%%%%%
\begin{equation}\label{eq:S1_pheno_size}
\begin{aligned}
\boldsymbol{\rm B)}&\qquad \frac{y_{1\,33}^\tau x_{1\,23}^{\tau\,*}}{M_{S_1^\tau}^2} \sim -0.5\, \text{TeV}^{-2}\,, \\ 
\boldsymbol{\rm A)}&\qquad \frac{y_{1\,33}^\tau (V_{cs} y_{1\,23}^{\tau\,*} + V_{cb} y_{1\,33}^{\tau\,*})}{M_{S_1^\tau}^2} \sim 0.2\, \text{TeV}^{-2}\,.
\end{aligned}
\end{equation}
%%%%%%%%%%%%%
A few considerations are now in order. Starting from the $C_{S_L}^\tau = -4 C_T^\tau$ scenario in Eq.~\eqref{eq:S1_bc} one can infer that, for couplings of order unity, the LQ mass is of order $M_{S_1^\tau} \sim 1.5\,\text{TeV}$. Even if $y_{1\,23}^\tau$ is now assumed to be vanishing, we nevertheless obtain a vectorial contribution $C_{V_L}^\tau \propto V_{cb} y_{1\,33}^\tau  y_{1\,33}^{\tau\,*}$, which is, however, negligible due to the CKM suppression: we are therefore consistent with scenario $\textbf{B)}$ of Eq.~\eqref{eq:bclnu_gf}, where $C_{V_L}^\tau$ is assumed to be 0.

If, on the other hand, one would like to pursue the vectorial solution identified by scenario $\textbf{A)}$ in Eq.~\eqref{eq:bclnu_gf}, a non-vanishing value for $y_{1\,23}^\tau$ is required together with a vanishing $x_{1\,23}^\tau$, in order to remove the scalar/tensor WCs while evading CKM suppression in the vectorial one. In this scenario, coupling of order unity would imply for the LQ a mass of order $m_{S_1^\tau} \sim 3\,\text{TeV}$. However, with this new choice of non-vanishing parameters a contribution for $C_{\nu\bar\nu}^\tau$ is implied as well, equal to $\sim -130$ and well above the current experimental bounds given at Eq.~\eqref{eq:bsnunu_gf}. Such a scenario would therefore require some additional mechanism in order to avoid the $B\to K^{(*)}\nu\bar\nu$ bounds, like e.g. the one proposed in Ref.~\cite{Crivellin:2017zlb}.

%%%%%%%%%%%%%%%%%%%%%%%%%%%%%%%%%

\subsubsection{Doublet Leptoquarks}
\label{sec:R2intro}

Moving on to weak doublets, the $R_2\equiv (\mathbf{3},\mathbf{2},7/6)$ scalar LQ Lagrangian is given by
%%%%%%%%%%%%%
\begin{equation}
    \label{eq:R2yukawaL}
    \mathcal{L}_{\mathcal{Y}_{R_2}} \,  = - y_{2\,ij}^a \Bar{u}_{R,\,i} R_2^{a,\,l} \epsilon^{lm} L_{L,\,j}^{m} + x_{2\,ij}^a \Bar{e}_{R,\,i} R_2^{a*} Q_{L,\,j} + \text{h.c.}\,,
\end{equation}
%%%%%%%%%%%%%
Due to $R_2$ being a doublet, it can either couple to a left-handed lepton doublet and a right-handed quark singlet, or vice-versa. These interactions are mediated by the arbitrary complex $3\times3$ matrices $y_{2}^a$ and $x_{2}^a$, respectively.

Similarly, the Lagrangian for $\tilde R_2\equiv (\mathbf{3},\mathbf{2},1/6)$ reads
%%%%%%%%%%%%%
\begin{equation}
    \label{eq:R2tyukawaL}
    \mathcal{L}_{\mathcal{Y}_{\tilde R_2}} \,  = - \tilde y_{2\,ij}^a \Bar{d}_{R,\,i} \tilde R_2^{a,\,l} \epsilon^{lm} L_{L,\,j}^{m} + \text{h.c.}\,.
\end{equation}
%%%%%%%%%%%%%
Analogously to Eq.~\eqref{eq:S1tyukawaL}, due to the different hypercharges of $R_2$ and $\tilde R_2$ only an interaction with a left-handed lepton doublet and a right-handed quark singlet is allowed for the latter, parameterized by the arbitrary complex $3\times3$ matrix $\tilde y_{2}^a$.

The doublet scalar $R_2^\tau$ LQ becomes phenomenologically relevant for us once the couplings $y_{2\,23}^\tau$ and $x_{2\,33}^{\tau}$ are allowed to be non-vanishing. Indeed, it contributes to $b\to c\tau\nu$ transitions via~\cite{Angelescu:2018tyl}
%%%%%%%%%%%%
\begin{equation}
C_{S_L}^\tau(\mu_{\rm LQ}) = 4 C_T^\tau(\mu_{\rm LQ}) = \frac{v^2}{4 V_{cb}}\frac{y_{2\,23}^\tau x_{2\,33}^{\tau\,*}}{M_{R_2^\tau}^2}\,,
\end{equation}
%%%%%%%%%%%%%
at the matching scale $\mu_{\rm LQ}=M_{R_2^\tau}\sim 2$ TeV. Once again, due to RGE effects the relation among the coefficients reads $C_{S_L}^\tau(\mu_b) = 8.4 C_T^\tau(\mu_b)$ at the low scale~\cite{Gonzalez-Alonso:2017iyc,Aebischer:2018acj}. The bound reported for scenario $\textbf{C)}$ in Eq.~\eqref{eq:bclnu_gf} can therefore be recast into a constraint on the parameter ratio 
%%%%%%%%%%%%
\begin{equation}\label{eq:R2_pheno_size}
\frac{y_{2\,23}^\tau x_{2\,33}^{\tau\,*}}{M_{R_2^\tau}^2} \sim 1.5\, \text{TeV}^{-2}\,,
\end{equation}
%%%%%%%%%%%%%
where we assumed one of the two coupling to be purely real and the other purely imaginary. Assuming for each coupling a size $\sim 1$ would imply a mass for the LQ below 1 TeV, already excluded by current constraints; it is however enough to require their size to be $\sim\sqrt{2}$, which is still below the current bounds obtained from searches for pair-produced LQs at the LHC, to obtain a mass of the order $M_{R_2^\tau} \sim 1.7\,\text{TeV}$, heavy enough to evade present limits. See Ref.~\cite{Angelescu:2021lln} and references therein for a detailed discussion on the matter.

%%%%%%%%%%%%%%%%%%%%%%%%%%%%%%%%%

\subsubsection{Triplet Leptoquarks}
\label{sec:S3intro}

We conclude this Section describing the interactions among the weak triplet $S_3\equiv (\mathbf{\Bar{3}},\mathbf{3},1/3)$ and the SM fields, ruled by the following Lagrangian:
%%%%%%%%%%%%%%
\begin{equation}
    \label{eq:S3yukawaL}
    \mathcal{L}_{\mathcal{Y}_{S_3}} \,  =  y_{3\,ij}^a \Bar{Q}^{C,\,l}_{L,\,i} \epsilon^{lm} ( \tau^k S_3^{a,\,k} )^{mn} L_{L,\,j}^{n} + \text{h.c.}\,,
\end{equation}
%%%%%%%%%%%%%
where $\tau^k$ are the Pauli matrices, with $k=1,2,3$. The contraction $( \tau^k S_3^{a,\,k} )$ can also be written as $( \Vec\tau \cdot \Vec S_3^a )$, as originally done in Ref.~\cite{Buchmuller:1986zs}. Due to its triplet nature, $S_3$ LQs can couple only with two left-handed SM fermions through the arbitrary complex $3\times3$ matrix $y_{3}^a$, analogously to the first term of Eq.~\eqref{eq:S1yukawaL}.

The triplet LQ has relevant phenomenological implications on $b\to s\ell^+\ell^-$ transitions. Indeed, allowing non-vanishing values for the couplings $y_{3\,3\ell}^\ell$ and $y_{3\,2\ell}^{\ell}$, with $\ell=e,\mu \equiv 1,2   $, it is possible to obtain contributions of the form~\cite{Angelescu:2018tyl}
%%%%%%%%%%%%
\begin{equation}\label{eq:S3_CL}
C_{L}^\ell(\mu_{\rm LQ}) = \frac{\pi v^2}{V_{tb} V_{ts}^*\alpha_{\rm em}}\frac{y_{3\,3\ell}^\ell y_{3\,2\ell}^{\ell\,*}}{M_{S_3^\ell}^2}\,,
\end{equation}
%%%%%%%%%%%%%
at the matching scale $\mu_{\rm LQ}=M_{S_3^\ell}$. Remembering that $C_9$ and $C_{10}$ do not run in QCD, the result for scenario $\textbf{II)}$ in Eq.~\eqref{eq:bsll_gf} can be directly applied, and implies for the NP parameter ratio the value
%%%%%%%%%%%%
\begin{equation}\label{eq:S3_pheno_size}
\frac{y_{3\,3\ell}^\ell y_{3\,2\ell}^{\ell\,*}}{m_{S_3^\ell}^2} \sim 0.001\, \text{TeV}^{-2}\,.
\end{equation}
%%%%%%%%%%%%%
Assuming the couplings to be of order unity, we can therefore infer the scale of the LQ mass to be $M_{S_3^\ell} \sim 30\,\text{TeV}$.

It is worth mentioning that, due to $SU(2)$ invariance, allowing additional couplings to $\tau$ would induce contributions to $b\to c\tau\nu$ transitions as well, similar to the ones obtained for $S_1$ LQs. However, the sign of such contributions would be strictly negative due to additional constraints coming, e.g., from $\Delta_{m_{B_s}}$~\cite{Angelescu:2018tyl} and hence not phenomenologically interesting, unless additional symmetries are imposed to the Lagrangian~\cite{Crivellin:2017zlb}. On the other hand, and again in a similar fashion to what is observed for the singlet LQ, contributions to $b \to s \nu\bar\nu$ transitions are unavoidable in this channel as well, and take the form
%%%%%%%%%%%%
\begin{equation}
C_{\nu\bar\nu}^\ell = \frac{\pi v^2}{V_{tb} V_{ts}^*\alpha_{\rm em}}\frac{y_{3\,3\ell}^\ell y_{3\,2\ell}^{\ell\,*}}{M_{S_3^\ell}^2}\,.
\end{equation}
%%%%%%%%%%%%%
In this scenario, however, the induced size on $C_{\nu\bar\nu}^\ell$ from $b\to s\ell^+\ell^-$ data would correspond to $C_{\nu\bar\nu}^\ell\sim -0.6$. This value is well within the current bounds, even when allowing for NP coupled to two lepton families which imply a  more stringent bound than the one given in Eq.~\eqref{eq:bsnunu_gf}.

%%%%%%%%%%%%%%%%%%%%%%%%%%%%%%%%%

\section{Renormalisation Group Equations}\label{sec:rge}
In this section we report the RGE of theories in which the SM sector is amended by an arbitrary number of Higgs doublets and scalar LQs. We start by giving in Sec.~\ref{sec:rge_gen} the RGEs for a fully generic theory with multiple copies of all the five scalar LQs. We then move to phenomenologically relevant cases, reporting the results obtained when the SM extended either with $(S_1,S_3)$ LQs or with $(R_2,S_3)$ LQs, in Sec.~\ref{sec:rge_s1s3} and Sec.~\ref{sec:rge_r2s3} respectively. All our results listed below correspond to the convention of our Lagrangians in Eqs.~\eqref{eq:yukawalagrangian}-\eqref{eq:S3yukawaL}. We give our results at the one-loop level of precision working in the $\overline{\rm MS}$-scheme, which we obtained adopting the findings of Refs.~\cite{Machacek:1983tz,Machacek:1983fi} to our specific scenarios. 

RGE effects of couplings in theories with LQs have been studied before: recently, the two-loop RGEs for couplings have been derived including the necessary one-loop threshold corrections for the gauge and SM Yukawa couplings~\cite{Banik:2023ogi} and used to study coupling unification at high scales. Similarly, in Ref.~\cite{Bandyopadhyay:2021kue} two-loop RGEs were studied to assess perturbative unitarity and vacuum stability up to the Planck scale. Conversely, the authors of Ref.~\cite{Kowalska:2020gie} have studied the implications of an ultraviolet (UV) fixed point at the Planck scale, motivated by asymptotically safe gravity, on low-energy Leptoquark couplings. The couplings have  been evolved to low energies and confronted with the flavour anomalies. While these papers employ RGEs to analyze UV properties of LQ theories, our study addresses the IR behaviour of LQ couplings. The RGE of gauge couplings was studied in Refs.~\cite{Bandyopadhyay:2021kue,Greljo:2022jac} to check for Landau poles at high energies.

%%%%%%%%%%%%%%%%%%%%%%%%%%%%%%%%%

\subsection{General Results}\label{sec:rge_gen}
Let us report here the RGE for the most general case, where arbitrary copies of the Higgs doublet and the five scalar LQs are allowed.

We start by giving the RGE for the gauge couplings $g_1$, $g_2$ and $g_3$, which we remember are connected to the $U(1)$, $SU(2)$ and $SU(3)$ gauge couplings by the convention $g_1\equiv\sqrt{3/5}g'$, $g_2\equiv g$ and $g_3\equiv g_s$. The RGE read
%%%%%%%%%%
\begin{align}
    16\pi^2\mu\frac{d}{d\mu}g_1 &= \,\, g_1^3 \Bigl( \frac{4}{3}n_{f} + \frac{1}{10} n_{H} + \frac{1}{15} n_{S_1} + \frac{16}{15} n_{\tilde S_1} + \frac{49}{30} n_{R_2} + \frac1{30} n_{\tilde R_2} + \frac{1}{5} n_{S_3} \Bigr)\,, \label{eq:RGEg1}\\[5pt]
%%%%%%%%%%
    16\pi^2\mu\frac{d}{d\mu}g_2 &= \,\, g_2^3\Bigl( -\frac{22}{3} + \frac{4}{3}n_{f} + \frac16 n_{H} + \frac12 n_{R_2} + \frac12 n_{\tilde R_2} + 2 n_{S_3}\Bigr)\,, \label{eq:RGEg2}\\[5pt]
%%%%%%%%%%
    16\pi^2\mu\frac{d}{d\mu}g_3 &= \,\, g_3^3 \Bigl(-11 + \frac{4}{3}n_{f} +  \frac{1}{6} n_{S_1} + \frac16 n_{\tilde S_1} + \frac{1}{3} n_{R_2} + \frac13 n_{\tilde R_2} + \frac{1}{2} n_{S_3}\Bigr)\,, \label{eq:RGEg3}
\end{align}
%%%%%%%%%%
where $n_f$ represents the number of SM flavours, $n_H$ is the number of scalar Higgs doublets, and $n_{S_1}$, $n_{\tilde S_1}$, $n_{R_2}$, $n_{\tilde R_2}$, $n_{S_3}$ are the numbers of  ${S_1}$, ${\tilde S_1}$, ${R_2}$, ${\tilde R_2}$, ${S_3}$ scalar LQs, respectively.

Before moving on to the RGE for the Yukawa couplings it is useful to define several quantities which will later allow us to state these RGE in a more compact and intuitive way. In particular, we give below the field renormalisation constants for all the relevant fields, namely the scalar Higgs doublets, the scalar LQs and the SM fermions.

\begin{figure}
  \includegraphics[width=\textwidth]{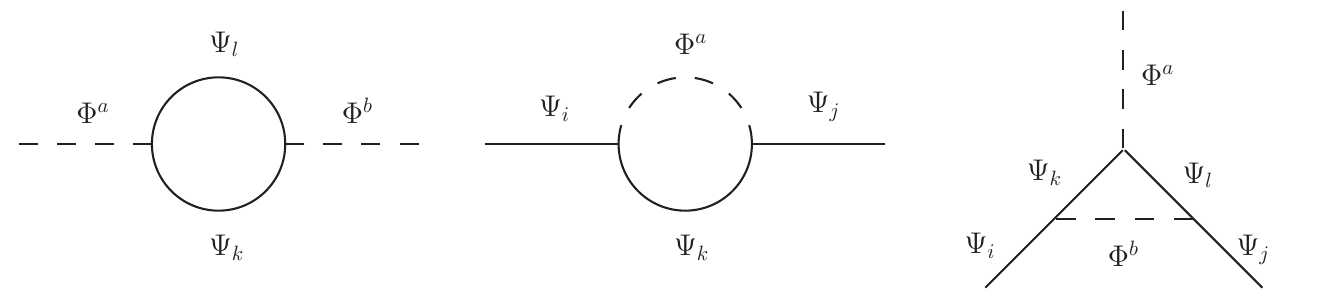}
  \caption{Diagrammatic representations of the contributions to the beta functions of the Yukawa couplings. $\Phi$ represents any scalar field, namely a Higgs doublet or a LQ, while $\Psi$ represents any SM fermion field.\label{fig:diags}
  }
\end{figure}

Starting from the six kind of scalars allowed in our theory, the contributions of the field renormalisation constants to the beta functions of the Yukawa couplings involve the following combinations of Yukawa matrices:
%%%%%%%%%%
\begin{eqnarray}
\begin{aligned}
    \label{eq:RGEscalarWF}
    \mathbf{T}^{ab} &=  \Tr\Bigl[ N_c {Y^a_u}^\dagger Y_u^b + N_c {Y^a_d}^\dagger Y_d^b +  {Y^a_e}^\dagger Y_e^b\Bigr]\,,\\
%%%%%%%%%%
    \mathbf{T}_{1}^{ab} &= \Tr \Bigl[ 2 y_1^{a} {y_1^b}^\dagger + x_1^{a} {x_1^b}^\dagger \Bigr]\,, \qquad
    \mathbf{T}_{2}^{ab} = \Tr \Bigl[ y_2^{a} {y_2^b}^\dagger + x_2^{a} {x_2^b}^\dagger \Bigr]\,,\\
%%%%%%%%%%
    \mathbf{\tilde T}_{1}^{ab} &= \Tr \Bigl[ \tilde x_1^{a} {\tilde x}_1^{b\dagger} \Bigr]\,, \qquad
    \mathbf{\tilde T}_{2}^{ab} = \Tr \Bigl[ \tilde y_2^{a} \tilde y_2^{b\dagger} \Bigr]\,, \qquad
    \mathbf{T}_{3}^{ab} = \Tr \Bigl[ 2 y_3^{a} {y_3^b}^\dagger \Bigr] \,,
\end{aligned}
\end{eqnarray}
%%%%%%%%%%
where $N_c=3$ is the colour number, and $a,b=1,\dots,n_\alpha$ with $\alpha \in \{H,S_1,\tilde S_1,R_2,\tilde R_2, S_3\}$ is an index denoting possible multiple copies of each scalar. All terms in Eq.~\eqref{eq:RGEscalarWF} stem from diagrams involving fermion loops, like the left one in Fig.~\ref{fig:diags}.

Concerning the field renormalisation constants of the SM fermion fields, we start with the contributions from loops with Higgs fields, which are
%%%%%%%%%%
\begin{eqnarray}
\begin{aligned}
    \label{eq:RGEHbubbles}
    [\mathcal{Y}_Q]_{ij} &= \frac12\lt[Y_u^a {Y^a_u}^\dagger + Y_d^a {Y_d^a}^\dagger\rt]_{ij}\,, \qquad
    [\mathcal{Y}_L]_{ij}  = \frac12\lt[Y_e^a {Y_e^a}^\dagger\rt]_{ij}\,, \\
    [\mathcal{Y}_u]_{ij} &= \lt[{Y_u^a}^\dagger Y_u^a\rt]_{ij}\,, \qquad
    [\mathcal{Y}_d]_{ij}  = \lt[{Y_d^a}^\dagger Y_d^a\rt]_{ij}\,, \qquad
    [\mathcal{Y}_e]_{ij}  = \lt[{Y_e^a}^\dagger Y_e^a\rt]_{ij}\,,
\end{aligned}
\end{eqnarray}
%%%%%%%%%%
where we denote the $(i,j)$ element of the matrix $M$ by $[M]_{ij}$. Here and below, we adopt the convention that repeated indices are implicitly summed over.

The contribution for the fermion field renormalisations due to the insertion of LQ in a loop read
%%%%%%%%%%
\begin{eqnarray}
\begin{aligned}
    \label{eq:RGELQbubbles}
    [\mathcal{Y}_1]_{ij} &= \frac12\lt[y_1^a {y_1^a}^\dagger\rt]_{ij}\,, \quad
    [\mathcal{\widehat Y}_1]_{ij} = \frac{N_c}2\lt[{y_1^a}^\dagger y_1^a\rt]_{ij}\,, \quad
    [\mathcal{X}_1]_{ij} = \frac12\lt[x_1^a {x_1^a}^\dagger\rt]_{ij}\,, \\
%%%%%%%%%%%%
    [\mathcal{X}_{\tilde 1}]_{ij} &= \frac12\lt[\tilde x_1^a \tilde x_1^{a\dagger} \rt]_{ij}\,, \quad
    [\mathcal{\widehat X}_{\tilde 1}]_{ij} = \frac{N_c}2\lt[\tilde x_1^{a\dagger} \tilde x_1^a\rt]_{ij}\,, \quad
    [\mathcal{\widehat X}_1]_{ij} = \frac{N_c}2\lt[{x_1^a}^\dagger x_1^a\rt]_{ij}\,, \\
%%%%%%%%%%%%
    [\mathcal{Y}_2]_{ij} &= \lt[y_2^a {y_2^a}^\dagger\rt]_{ij}\,, \quad  
    [\mathcal{\widehat Y}_2]_{ij} = \frac{N_c}2\lt[{y_2^a}^\dagger y_2^a\rt]_{ij}\,, \quad
    [\mathcal{X}_2]_{ij}  = N_c\lt[x_2^a {x_2^a}^\dagger\rt]_{ij}\,, \\
%%%%%%%%%%%%
    [\mathcal{Y}_{\tilde 2}]_{ij} &= \lt[\tilde y_2^a \tilde y_2^{a\dagger}\rt]_{ij}\,, \quad
    [\mathcal{\widehat Y}_{\tilde 2}]_{ij} = \frac{N_c}2\lt[\tilde y_2^{a\dagger} \tilde y_2^a\rt]_{ij}\,, \quad
    [\mathcal{\widehat X}_2]_{ij} = \frac12\lt[{x_2^a}^\dagger x_2^a\rt]_{ij}\,, \\
%%%%%%%%%%%%
    [\mathcal{Y}_3]_{ij} &= \frac32\lt[y_3^a {y_3^a}^\dagger\rt]_{ij}\,, \quad
    [\mathcal{\widehat Y}_3]_{ij}  = \frac{3N_c}2\lt[{y_3^a}^\dagger y_3^a\rt]_{ij}\,.
\end{aligned}
\end{eqnarray}
%%%%%%%%%%

Combining Eq.~\eqref{eq:RGEHbubbles} and Eq.~\eqref{eq:RGELQbubbles}, both stemming from diagrams involving a fermion and a scalar in a loop as depicted in the center of Fig.~\ref{fig:diags}, allows us to finally define the total contribution to the field renormalisations of the SM fermions, which read
%%%%%%%%%%
\begin{eqnarray}
\begin{aligned}
    \label{eq:RGEfermionWF}
    [\mathcal{Y}_{QQ}]_{ij} &= [\mathcal{Y}_Q + \mathcal{Y}_1^* + \mathcal{\widehat X}_2 + \mathcal{Y}_3^*]_{ij} \,,
    \qquad
    [\mathcal{Y}_{LL}]_{ij} = [\mathcal{Y}_L + \mathcal{\widehat Y}_1 + \mathcal{\widehat Y}_2 + \mathcal{\widehat Y}_{\tilde 2} + \mathcal{\widehat Y}_3]_{ij}\,,
    \\
    [\mathcal{Y}_{uu}]_{ij} &= [\mathcal{Y}_u + \mathcal{X}_1^* + \mathcal{Y}_2]_{ij}\,,
    \qquad\qquad\quad\,\,
    [\mathcal{Y}_{dd}]_{ij} = [\mathcal{Y}_d + \mathcal{X}_{\tilde 1}^* + \mathcal{Y}_{\tilde 2}]_{ij}\,,
    \\
    [\mathcal{Y}_{ee}]_{ij} &= [\mathcal{Y}_e + \mathcal{\widehat X}_1 + \mathcal{\widehat X}_{\tilde 1} + \mathcal{X}_2]_{ij}\,,
\end{aligned}
\end{eqnarray}
%%%%%%%%%%
where the labels refer to the external fields.

Employing Eq.~\eqref{eq:RGEscalarWF} and Eq.~\eqref{eq:RGEfermionWF}, complemented by additional contributions from vertex corrections as the one shown in the right side of Fig.~\ref{fig:diags}, we are now ready to give the RGE for the Yukawa couplings introduced in Sec.~\ref{sec:lag}. The RGE of the SM Yukawa couplings defined in Eq.~\eqref{eq:yukawalagrangian} read
%%%%%%%%%%
\begin{align} 
    16\pi^2\mu\frac{d}{d\mu} [Y^a_u]_{ij} = \,\,  & [Y^a_u]_{ij} \Bigl( - 8 g_3^2 - \frac{9}{4} g_2^2 - \frac{17}{20} g_1^2 \Bigr) + {\mathbf{T}^{ab}}^* \, [Y^b_u]_{ij} + [\mathcal{Y}_{QQ}]_{ik} [Y^a_u]_{kj} + [Y^a_u]_{ik} [\mathcal{Y}_{uu}]_{kj} \nonumber\\
    &
    - 2 \left([Y_{d}^b {Y^a_d}^\dagger Y_{u}^b]_{ij} - [y_1^b Y_{e}^{a} {x_1^b}^\dagger]^*_{ij} + [y_2^b Y_e^a x_2^b]^\dagger_{ij}\right) \,, \label{eq:RGEYu} \\[5pt]
%%%%%%%%%%%%
    16\pi^2\mu\frac{d}{d\mu} [Y^a_d]_{ij} = \,\, & [Y^a_d]_{ij} \Bigl( - 8 g_3^2 - \frac{9}{4} g_2^2 - \frac{1}{4} g_1^2 \Bigr) + \mathbf{T}^{ab} \, [Y^b_d]_{ij} + [\mathcal{Y}_{QQ}]_{ik} [Y^a_d]_{kj} + [Y^a_d]_{ik} [\mathcal{Y}_{dd}]_{kj} \nonumber\\
    &
    - 2 [Y_{u}^b {Y^a_u}^\dagger Y_{d}^b]_{ij} \,, \label{eq:RGEYd} \\[5pt]
%%%%%%%%%%%%%
    16\pi^2\mu\frac{d}{d\mu} [Y^a_e]_{ij} = \,\, & [Y^a_e]_{ij} \Bigl(  - \frac{9}{4} g_2^2 - \frac{9}{4} g_1^2 \Bigr) +  \mathbf{T}^{ab} \, [Y^b_e]_{ij} + [\mathcal{Y}_{LL}]_{ik} [Y^a_e]_{kj} + [Y^a_e]_{ik} [\mathcal{Y}_{ee}]_{kj} \nonumber\\
    &
    + 2N_c \left([{y_1^b}^\dagger {Y_u^a}^* x_1^b]_{ij} - [x_2^b Y_u^a y_2^b]^\dagger_{ij} \right) \,. \label{eq:RGEYe}
\end{align}
%%%%%%%%%%

The RGE of the singlet LQs Yukawa couplings defined in Eqs.~\eqref{eq:S1yukawaL}-\eqref{eq:S1tyukawaL} read
%%%%%%%%%%
\begin{align}
    16\pi^2\mu\frac{d}{d\mu} [y^a_1]_{ij} = \,\, &
    [y^a_1]_{ij} \Bigl( - 4 g_3^2 - \frac{9}{2} g_2^2 - \frac{1}{2} g_1^2 \Bigr) + \mathbf{T}_{1}^{ab} [y^b_1]_{ij} + [\mathcal{Y}_{QQ}]^*_{ik} [y^a_1]_{kj} + [y^a_1]_{ik} [\mathcal{Y}_{LL}]_{kj} \nonumber\\
    &
    + 2 \left([{Y_u^b}^* x_1^a {Y_e^b}^\dagger]_{ij} - [{x_2^b}^T {x_1^a}^T y_2^b]_{ij} \right) \,, \label{eq:RGEy1} \\[5pt]
%%%%%%%%%%%%
    16\pi^2\mu\frac{d}{d\mu} [x^a_1]_{ij} = \,\, & [x^a_1]_{ij} \Bigl( - 4 g_3^2 - \frac{13}{5} g_1^2 \Bigr) +  \mathbf{T}_{1}^{ab} [x^b_1]_{ij} + [\mathcal{Y}_{uu}]^*_{ik} [x^a_1]_{kj} + [x^a_1]_{ik} [\mathcal{Y}_{ee}]_{kj} \nonumber\\
    &
    + 4 \left([{Y_u^b}^T y_1^a Y_e^b]_{ij} - [{y_2^b}^* {y_1^a}^T {x_2^b}^\dagger]_{ij}\right) \,, \label{eq:RGEx1} \\[5pt]
%%%%%%%%%%%%
    16\pi^2\mu\frac{d}{d\mu} [\tilde x^a_1]_{ij} = \,\, & [\tilde x^a_1]_{ij} \Bigl( - 4 g_3^2 - 2 g_1^2 \Bigr) +  \mathbf{\tilde T}_{1}^{ab} [\tilde x^b_1]_{ij} + [\mathcal{Y}_{dd}]^*_{ik} [\tilde x^a_1]_{kj} + [\tilde x^a_1]_{ik} [\mathcal{Y}_{ee}]_{kj} \,. \label{eq:RGEtx1}
\end{align}
%%%%%%%%%%

The RGE of the doublet LQs Yukawa couplings defined in Eqs.~\eqref{eq:R2yukawaL}-\eqref{eq:R2tyukawaL} read
\begin{align}
    16\pi^2\mu\frac{d}{d\mu} [y^a_2]_{ij} = \,\, & [y^a_2]_{ij} \Bigl( - 4 g_3^2 - \frac{9}{4} g_2^2 - \frac{5}{4} g_1^2 \Bigr) +  \mathbf{T}_{2}^{ab} [y^b_2]_{ij} + [\mathcal{Y}_{uu}]_{ik} [y^a_2]_{kj} + [y^a_2]_{ik} [\mathcal{Y}_{LL}]_{kj} \nonumber\\
    &
    - 2 \left([Y_e^b x_2^a Y_u^b]^\dagger_{ij} + [{x_1^b}^* {x_2^a}^* y_1^b]_{ij} \right) \,, \label{eq:RGEy2} \\[5pt]
%%%%%%%%%%%%%%
    16\pi^2\mu\frac{d}{d\mu} [x^a_2]_{ij} = \,\, & [x^a_2]_{ij} \Bigl( - 4 g_3^2 - \frac{9}{4} g_2^2 - \frac{37}{20} g_1^2 \Bigr) +  \mathbf{T}_{2}^{ab} [x^b_2]_{ij} + [\mathcal{Y}_{ee}]_{ik} [x^a_2]_{kj} + [x^a_2]_{ik} [\mathcal{Y}_{QQ}]_{kj} \nonumber\\
    &
    - 2 \left([Y_u^b y_2^a Y_e^b]^\dagger_{ij} + [{y_1^b}^* {y_2^a}^T x_1^b]^\dagger_{ij}\right) \,, \label{eq:RGEx2} \\[5pt]
%%%%%%%%%%%%
    16\pi^2\mu\frac{d}{d\mu} [\tilde y^a_2]_{ij} = \,\, & [\tilde y^a_2]_{ij} \Bigl( - 4 g_3^2 - \frac94 g_2^2 - \frac{13}{20} g_1^2 \Bigr) +  \mathbf{\tilde T}_{2}^{ab} [\tilde y^b_2]_{ij} + [\mathcal{Y}_{dd}]_{ik} [\tilde y^a_2]_{kj} + [\tilde y^a_2]_{ik} [\mathcal{Y}_{LL}]_{kj} \,. \label{eq:RGEty2}
\end{align}
%%%%%%%%%%

Finally, the RGE of the triplet LQ Yukawa coupling defined in Eq.~\eqref{eq:S3yukawaL} reads
\begin{align}
    16\pi^2\mu\frac{d}{d\mu} [y^a_3]_{ij} = \,\, & [y^a_3]_{ij} \Bigl( - 4 g_3^2 -  \frac{9}{2} g_2^2 - \frac{1}{2} g_1^2 \Bigr) +  \mathbf{T}_{3}^{ab} [y^b_3]_{ij} + [\mathcal{Y}_{QQ}]^*_{ik} [y^a_3]_{kj} + [y^a_3]_{ik} [\mathcal{Y}_{LL}]_{kj} \,. \label{eq:RGEy3}
\end{align}

%%%%%%%%%%%%%%%%%%%%%%%%%%%%%%%%%

\subsection{The SM Extended by \texorpdfstring{$\mathbf{S_1}$}{S1} and \texorpdfstring{$\mathbf{S_3}$}{S3} LQs}\label{sec:rge_s1s3}
Let us now move our focus to the first of the two phenomenologically relevant models, whose RGE implications will be studied in Sec.~\ref{sec:fixedpoints}, namely the one consisting in the extension of the SM with $S_1$ and $S_3$ scalar LQs, and no additional Higgs doublets. This kind of models has been originally proposed in Ref.~\cite{Crivellin:2017zlb} and subsequently embedded in a composite Higgs model in Ref.~\cite{Marzocca:2018wcf}. They originally proposed a singlet LQ $S_1$ to account for the anomalies in $b\to c\tau\nu$ transitions, and a triplet LQ $S_3$ for addressing data in $b\to s\mu\mu$ decays, as shown in Fig.~\ref{fig:bdec}. As detailed in Sec.~\ref{sec:Leptoquarks}, the requirement of lepton flavour universality in $b\to s\ell^+\ell^-$ transitions implies now the presence of multiple copies of $S_3$ LQs. While a similar behaviour is not required for the $S_1$ LQ, we will however maintain a degree of generality here and allow for multiple copies of this scalar LQ as well.

For this kind of theory Eqs.~\eqref{eq:RGEg1}-\eqref{eq:RGEg3} condense to
%%%%%%%%%%
\begin{align}
    16\pi^2\mu\frac{d}{d\mu}g_1 &= \,\, g_1^3 \Bigl( \frac{4}{3}n_{f} + \frac{1}{10} + \frac{1}{15} n_{S_1} + \frac{1}{5} n_{S_3} \Bigr)\,, \label{eq:RGEg1_s1s3}\\[5pt]
%%%%%%%%%%
    16\pi^2\mu\frac{d}{d\mu}g_2 &= \,\, g_2^3\Bigl( -\frac{22}{3} + \frac{4}{3}n_{f} + \frac{1}{6} + 2 n_{S_3}\Bigr)\,, \label{eq:RGEg2_s1s3}\\[5pt]
%%%%%%%%%%
    16\pi^2\mu\frac{d}{d\mu}g_3 &= \,\, g_3^3 \Bigl(-11 + \frac{4}{3}n_{f} +  \frac{1}{6} n_{S_1} + \frac{1}{2} n_{S_3}\Bigr)\,. \label{eq:RGEg3_s1s3}
\end{align}
%%%%%%%%%%

The contribution from scalar field renormalisations are found from Eq.~\eqref{eq:RGEscalarWF}, by specifying to only one Higgs doublet and SM Yukawas $Y^a_{u,d,e}\equiv Y_{u,d,e}$, and hence $\mathbf{T}^{ab}\equiv \mathbf{T}$. The fermion field renormalisations of Eq.~\eqref{eq:RGEfermionWF} are altered by the reduced scalar sector of the theory, and now read
%%%%%%%%%%
\begin{eqnarray}
\begin{aligned}
    \label{eq:RGEfermionWF_s1s3}
    [\mathcal{Y}_{QQ}^{\, '}]_{ij} &= [\mathcal{Y}_Q + \mathcal{Y}_1^* + \mathcal{Y}_3^*]_{ij} \,,
    \qquad
    [\mathcal{Y}_{LL}^{\, '}]_{ij} = [\mathcal{Y}_L + \mathcal{\widehat Y}_1 + \mathcal{\widehat Y}_3]_{ij}\,,
    \\
    [\mathcal{Y}_{uu}^{\, '}]_{ij} &= [\mathcal{Y}_u + \mathcal{X}_1^*]_{ij}\,,
    \quad\quad\quad\quad\;\;\,
    [\mathcal{Y}_{dd}^{\, '}]_{ij} = [\mathcal{Y}_d]_{ij}\,,
    \\
    [\mathcal{Y}_{ee}^{\, '}]_{ij} &= [\mathcal{Y}_e + \mathcal{\widehat X}_1]_{ij}\,.
\end{aligned}
\end{eqnarray}
%%%%%%%%%%

We have now all the ingredients necessary to give the RGE for the Yukawa couplings of an extension of the SM by multiple copies of $S_1$ and $S_3$ LQs. The RGE of the SM Yukawa couplings defined in Eq.~\eqref{eq:yukawalagrangian} read
%%%%%%%%%%
\begin{align} 
    16\pi^2\mu\frac{d}{d\mu} [Y_u]_{ij} = \,\,  & [Y_u]_{ij} \Bigl( - 8 g_3^2 - \frac{9}{4} g_2^2 - \frac{17}{20} g_1^2 \Bigr) + {\mathbf{T}}^* \, [Y_u]_{ij} + [\mathcal{Y}_{QQ}^{\, '}]_{ik} [Y_u]_{kj} + [Y_u]_{ik} [\mathcal{Y}_{uu}^{\, '}]_{kj} \nonumber\\
    &
    - 2 \left([Y_{d} Y_d^\dagger Y_{u}]_{ij} - [y_1^b Y_{e}{x_1^b}^\dagger]^*_{ij}\right) \,, \label{eq:RGEYu_s1s3} \\[5pt]
%%%%%%%%%%%%
    16\pi^2\mu\frac{d}{d\mu} [Y_d]_{ij} = \,\, & [Y_d]_{ij} \Bigl( - 8 g_3^2 - \frac{9}{4} g_2^2 - \frac{1}{4} g_1^2 \Bigr) + \mathbf{T} \, [Y_d]_{ij} + [\mathcal{Y}_{QQ}^{\, '}]_{ik} [Y_d]_{kj} + [Y_d]_{ik} [\mathcal{Y}_{dd}^{\, '}]_{kj} \nonumber\\
    &
    - 2 [Y_{u} {Y_u}^\dagger Y_{d}]_{ij} \,, \label{eq:RGEYd_s1s3} \\[5pt]
%%%%%%%%%%%%%
    16\pi^2\mu\frac{d}{d\mu} [Y_e]_{ij} = \,\, & [Y_e]_{ij} \Bigl(  - \frac{9}{4} g_2^2 - \frac{9}{4} g_1^2 \Bigr) +  \mathbf{T} \, [Y_e]_{ij} + [\mathcal{Y}_{LL}^{\, '}]_{ik} [Y_e]_{kj} + [Y_e]_{ik} [\mathcal{Y}_{ee}^{\, '}]_{kj} \nonumber\\
    &
    + 2N_c [{y_1^b}^\dagger Y_u^* x_1^b]_{ij} \,. \label{eq:RGEYe_s1s3}
\end{align}
%%%%%%%%%%

The RGE of the singlet LQs Yukawa couplings defined in Eq.~\eqref{eq:S1yukawaL} read
%%%%%%%%%%
\begin{align}
    16\pi^2\mu\frac{d}{d\mu} [y^a_1]_{ij} = \,\, &
    [y^a_1]_{ij} \Bigl( - 4 g_3^2 - \frac{9}{2} g_2^2 - \frac{1}{2} g_1^2 \Bigr) + \mathbf{T}_{1}^{ab} [y^b_1]_{ij} + [\mathcal{Y}_{QQ}^{\, '}]^*_{ik} [y^a_1]_{kj} + [y^a_1]_{ik} [\mathcal{Y}_{LL}^{\, '}]_{kj} \nonumber\\
    &
    + 2 [Y_u^* x_1^a Y_e^\dagger]_{ij} \,, \label{eq:RGEy1_s1s3} \\[5pt]
%%%%%%%%%%%%
    16\pi^2\mu\frac{d}{d\mu} [x^a_1]_{ij} = \,\, & [x^a_1]_{ij} \Bigl( - 4 g_3^2 - \frac{13}{5} g_1^2 \Bigr) +  \mathbf{T}_{1}^{ab} [x^b_1]_{ij} + [\mathcal{Y}_{uu}^{\, '}]^*_{ik} [x^a_1]_{kj} + [x^a_1]_{ik} [\mathcal{Y}_{ee}^{\, '}]_{kj} \nonumber\\
    &
    + 4 [Y_u^T y_1^a Y_e]_{ij} \,. \label{eq:RGEx1_s1s3}
\end{align}
%%%%%%%%%%

Finally, the RGE of the triplet LQ Yukawa coupling defined in Eq.~\eqref{eq:S3yukawaL} reads
\begin{align}
    16\pi^2\mu\frac{d}{d\mu} [y^a_3]_{ij} = \,\, & [y^a_3]_{ij} \Bigl( - 4 g_3^2 -  \frac{9}{2} g_2^2 - \frac{1}{2} g_1^2 \Bigr) +  \mathbf{T}_{3}^{ab} [y^b_3]_{ij} + [\mathcal{Y}_{QQ}^{\, '}]^*_{ik} [y^a_3]_{kj} + [y^a_3]_{ik} [\mathcal{Y}_{LL}^{\, '}]_{kj} \,. \label{eq:RGEy3_s1s3}
\end{align}

%%%%%%%%%%%%%%%%%%%%%%%%%%%%%%%%%

\subsection{The SM Extended by \texorpdfstring{$\mathbf{R_2}$}{R2} and \texorpdfstring{$\mathbf{S_3}$}{S3} LQs}\label{sec:rge_r2s3}
The second phenomenologically relevant model consists of the extension of the SM with $R_2$ and $S_3$ scalar LQs, and again no additional Higgs doublets. This model was originally proposed in Ref.~\cite{Becirevic:2018afm} where the two LQs were embedded in an $SU(5)$ Grand Unification Theory (GUT) and, as shown in Fig.~\ref{fig:bdec}, employs the $R_2$ LQ to address data in $b\to c\tau\bar\nu$ decays, again in combination with the $S_3$ LQ to explain anomalies in $b\to s\ell^+\ell^-$ transitions. Similarly to the previous case, we will permit multiple copies for both scalar LQs.

For this kind of model the gauge coupling RGE from Eqs.~\eqref{eq:RGEg1}-\eqref{eq:RGEg3} condense to
%%%%%%%%%%
\begin{align}
    16\pi^2\mu\frac{d}{d\mu}g_1 &= \,\, g_1^3 \Bigl( \frac{4}{3}n_{f} + \frac{1}{10} + \frac{49}{30} n_{R_2} + \frac{1}{5} n_{S_3} \Bigr)\,, \label{eq:RGEg1_r2s3}\\[5pt]
%%%%%%%%%%
    16\pi^2\mu\frac{d}{d\mu}g_2 &= \,\, g_2^3\Bigl( -\frac{22}{3} + \frac{4}{3}n_{f} + \frac16 + \frac12 n_{R_2} + 2 n_{S_3}\Bigr)\,, \label{eq:RGEg2_r2s3}\\[5pt]
%%%%%%%%%%
    16\pi^2\mu\frac{d}{d\mu}g_3 &= \,\, g_3^3 \Bigl(-11 + \frac{4}{3}n_{f} +  \frac{1}{3} n_{R_2} + \frac{1}{2} n_{S_3}\Bigr)\,, \label{eq:RGEg3_r2s3}
\end{align}
%%%%%%%%%%

In a similar fashion to the previous scenario, the scalar field renormalisations are analogous to the ones given at Eq.~\eqref{eq:RGEscalarWF} specified to a single Higgs doublet, while for the fermion ones we now have
%%%%%%%%%%
\begin{eqnarray}
\begin{aligned}
    \label{eq:RGEfermionWF_r2s3}
    [\mathcal{Y}_{QQ}^{\, ''}]_{ij} &= [\mathcal{Y}_Q  + \mathcal{\widehat X}_2 + \mathcal{Y}_3^*]_{ij} \,,
    \qquad
    [\mathcal{Y}_{LL}^{\, ''}]_{ij} = [\mathcal{Y}_L + \mathcal{\widehat Y}_2 + \mathcal{\widehat Y}_3]_{ij}\,,
    \\
    [\mathcal{Y}_{uu}^{\, ''}]_{ij} &= [\mathcal{Y}_u + \mathcal{Y}_2]_{ij}\,,
    \quad\quad\quad\quad\;\;\;
    [\mathcal{Y}_{dd}^{\, ''}]_{ij} = [\mathcal{Y}_d]_{ij}\,,
    \\
    [\mathcal{Y}_{ee}^{\, ''}]_{ij} &= [\mathcal{Y}_e  + \mathcal{X}_2]_{ij}\,.
\end{aligned}
\end{eqnarray}
%%%%%%%%%%

We can now move on to the RGE equations for the Yukawa couplings in this kind of theory. The RGE of the SM Yukawa couplings defined in Eq.~\eqref{eq:yukawalagrangian} read
%%%%%%%%%%
\begin{align} 
    16\pi^2\mu\frac{d}{d\mu} [Y_u]_{ij} = \,\,  & [Y_u]_{ij} \Bigl( - 8 g_3^2 - \frac{9}{4} g_2^2 - \frac{17}{20} g_1^2 \Bigr) + \mathbf{T}^* \, [Y_u]_{ij} + [\mathcal{Y}_{QQ}^{\, ''}]_{ik} [Y_u]_{kj} + [Y_u]_{ik} [\mathcal{Y}_{uu}^{\, ''}]_{kj} \nonumber\\
    &
    - 2 \left([Y_{d} Y_d^\dagger Y_{u}]_{ij} + [y_2^b Y_e x_2^b]^\dagger_{ij}\right) \,, \label{eq:RGEYu_r2s3} \\[5pt]
%%%%%%%%%%%%
    16\pi^2\mu\frac{d}{d\mu} [Y_d]_{ij} = \,\, & [Y_d]_{ij} \Bigl( - 8 g_3^2 - \frac{9}{4} g_2^2 - \frac{1}{4} g_1^2 \Bigr) + \mathbf{T} \, [Y_d]_{ij} + [\mathcal{Y}_{QQ}^{\, ''}]_{ik} [Y_d]_{kj} + [Y_d]_{ik} [\mathcal{Y}_{dd}^{\, ''}]_{kj} \nonumber\\
    &
    - 2 [Y_{u} Y_u^\dagger Y_{d}]_{ij} \,, \label{eq:RGEYd_r2s3} \\[5pt]
%%%%%%%%%%%%%
    16\pi^2\mu\frac{d}{d\mu} [Y_e]_{ij} = \,\, & [Y_e]_{ij} \Bigl(  - \frac{9}{4} g_2^2 - \frac{9}{4} g_1^2 \Bigr) +  \mathbf{T} \, [Y_e]_{ij} + [\mathcal{Y}_{LL}^{\, ''}]_{ik} [Y_e]_{kj} + [Y_e]_{ik} [\mathcal{Y}_{ee}^{\, ''}]_{kj} \nonumber\\
    &
    - 2N_c [x_2^b Y_u y_2^b]^\dagger_{ij} \,. \label{eq:RGEYe_r2s3}
\end{align}
%%%%%%%%%%

The RGE of the doublet LQ Yukawa couplings defined in Eq.~\eqref{eq:R2yukawaL} read
%%%%%%%%%%%
\begin{align}
    16\pi^2\mu\frac{d}{d\mu} [y^a_2]_{ij} = \,\, & [y^a_2]_{ij} \Bigl( - 4 g_3^2 - \frac{9}{4} g_2^2 - \frac{5}{4} g_1^2 \Bigr) +  \mathbf{T}_{2}^{ab} [y^b_2]_{ij} + [\mathcal{Y}_{uu}^{\, ''}]_{ik} [y^a_2]_{kj} + [y^a_2]_{ik} [\mathcal{Y}_{LL}^{\, ''}]_{kj} \nonumber\\
    &
    - 2 [Y_e x_2^a Y_u]^\dagger_{ij} \,, \label{eq:RGEy2_r2s3} \\[5pt]
%%%%%%%%%%%
    16\pi^2\mu\frac{d}{d\mu} [x^a_2]_{ij} = \,\, & [x^a_2]_{ij} \Bigl( - 4 g_3^2 - \frac{9}{4} g_2^2 - \frac{37}{20} g_1^2 \Bigr) +  \mathbf{T}_{2}^{ab} [x^b_2]_{ij} + [\mathcal{Y}_{ee}^{\, ''}]_{ik} [x^a_2]_{kj} + [x^a_2]_{ik} [\mathcal{Y}_{QQ}^{\, ''}]_{kj} \nonumber\\
    &
    - 2 [Y_u y_2^a Y_e]^\dagger_{ij} \,. \label{eq:RGEx2_r2s3}
\end{align}
%%%%%%%%%%

Finally, the RGE of the triplet LQ Yukawa coupling defined in Eq.~\eqref{eq:S3yukawaL} reads
\begin{align}
    16\pi^2\mu\frac{d}{d\mu} [y^a_3]_{ij} = \,\, & [y^a_3]_{ij} \Bigl( - 4 g_3^2 -  \frac{9}{2} g_2^2 - \frac{1}{2} g_1^2 \Bigr) +  \mathbf{T}_{3}^{ab} [y^b_3]_{ij} + [\mathcal{Y}_{QQ}^{\, ''}]^*_{ik} [y^a_3]_{kj} + [y^a_3]_{ik} [\mathcal{Y}_{LL}^{\, ''}]_{kj} \,. \label{eq:RGEy3_r2s3}
\end{align}

%%%%%%%%%%%%%%%%%%%%%%%%%%%%%%%%%

\section{Phenomenology of Fixed Point Solutions}\label{sec:fixedpoints}
We have now collected all the necessary ingredients to perform the study of the RGE IR FPs, and to discuss their potential phenomenological implications for the BSM scenarios selected in Sec.~\ref{sec:rge_s1s3} and~\ref{sec:rge_r2s3}. Our aim is the investigation of solutions to the anomalies in $b\to s$ and $b \to c$ transitions with the IR FP values for such couplings.

As anticipated above, we will perform our studies in two distinct scenarios, differentiated by whether the SM sector is extended by (potentially multiple copies of) $S_1$ and $S_3$ LQs, or $R_2$ and $S_3$ LQs, respectively. In both scenarios we will first study the minimal case, where only one new field involved in $b \to c$ transitions is considered, namely either $S_1^\tau$ or $R_2^\tau$, while two new fields are allowed in the $b \to s$ sector due to the requirement of a LFU phenomenology, namely $S_3^e$ and $S_3^\mu$. Subsequently, we will also consider the case where 6 NP fields are included in the theory, i.e. three new fields connected to $b \to c$ transitions, namely either $S_1^e$, $S_1^\mu$ and $S_1^\tau$, or $R_2^e$, $R_2^\mu$ and $R_2^\tau$, and three new fields connected to the $b \to s$ sector, namely $S_3^e$, $S_3^\mu$ and $S_3^\tau$. 

To obtain the FP values for the couplings investigated below, we will employ the following procedure. As a starting point, for each specific SM extension we identify the minimal set of $n$ LQ couplings required to address anomalies in $b\to s$ and $b\to c$ transitions; therefore, we take their corresponding $n$ beta functions, which are given for the $(S_1,S_3)$ extension in Eqs.~\eqref{eq:RGEy1_s1s3}-\eqref{eq:RGEy3_s1s3} and for the $(R_2,S_3)$ one in Eqs.~\eqref{eq:RGEy2_r2s3}-\eqref{eq:RGEy3_r2s3}, and we consider them in the limit where all the other NP couplings vanish. Finally, we set all the SM couplings entering these beta functions to their experimental values evolved to the scale of 10 TeV,
which we choose as the low-energy scale of the RG evolution. We obtain in this way a set of $n$ equations (the beta functions of the LQ couplings of interest are equal to zero) in $n$ variables (the LQ couplings themselves), and the FP values for these couplings are therefore obtained requiring that all equations are fulfilled simultaneously.

Given the non linearity of the system and its high dimensionality, listing all the found solutions goes beyond the scope of our analysis. We will therefore restrict ourselves to reporting phenomenologically interesting FP solutions, namely those that comply with at least one of the following requirements:
\begin{itemize}
    \item[\emph{i})] all FP values for the couplings have to be non-vanishing;
    \item[\emph{ii})] the $S_3^e$ and $S_3^\mu$ couplings have to obey the relation $y_{3\, 21}^{e} y_{3\, 31}^{e} = y_{3\, 22}^{\mu} y_{3\, 32}^{\mu}$, required by the LFU scenario $\textbf{II)}$ in Eq.~\eqref{eq:bsll_gf};
    \item[\emph{iii})] if present, the product of the $R_2^\tau$ couplings $y_{2\,23}^\tau x_{2\,33}^{\tau\,*}$ has to be purely imaginary, in accordance with scenario $\textbf{C)}$ in Eq.~\eqref{eq:bclnu_gf}.
\end{itemize}

Once the couplings are determined by their FP values, the experimental constraints from the anomalies fix the values of the (squared) LQ masses. We will face two possible outcomes: \emph{a)} the FP values for the couplings are large enough to reproduce the desired phenomenology with sufficiently heavy LQ masses, not currently excluded by direct searches at collider. This will also allow us to give a prediction for $M_{\rm LQ}$, in the case where the low-energy physics is described by the FP values of the LQ couplings; or \emph{b)} the FP values are not large enough to explain the desired phenomenology, because the LQ are too light to comply with direct searches results.\footnote{A third possibility would consist to ascribe to LQs only a part of the NP contributions required to address the current experimental picture. This scenario would however require to further extend the NP sector to fully explain data, with a consequent modification of the RGE due to the presence of additional degrees of freedom. Such a scenario goes beyond the scope of this paper.} Nevertheless, also in this scenario useful conclusions can be drawn: indeed, it will imply that in order to explain $b\to s$ and $b \to c$ data, the values for (some of) the LQ couplings is required to be above the FP value. It is therefore interesting to estimate the scale where the Landau pole is induced by such a choice, since this scale can be interpreted as the upper bound for $M_{\rm QLU}$. We finally remark that for all scenarios discussed below we verified the absence of low-scale Landau poles in the gauge couplings, in agreement with the findings of Refs.~\cite{Bandyopadhyay:2021kue,Greljo:2022jac}.

%%%%%%%%%%%%%%%%%%%%%%%%%%%%%%%%%

\subsection{The \texorpdfstring{$\mathbf{(S_1,S_3)}$}{(S1,S3)} Extension}\label{sec:fp_s1s3}
We start our analysis from the scenario where the SM is extended by one copy of the singlet LQ, $S_1^\tau$, and two copies of the triplet one, $S_3^e$ and $S_3^\mu$. Indeed, as detailed in Sec.~\ref{sec:S1intro} and~\ref{sec:S3intro} respectively, $S_1^\tau$ is capable to reproduce the desired low-scale phenomenology for $b\to c\tau\bar\nu$ decays once the couplings $y_{1\,33}^\tau$ and $x_{1\,23}^\tau$ are non-vanishing, while $S_3^e$ and $S_3^\mu$ can produce the correct low-energy effect in $b\to s\ell^+\ell^-$ transitions when the couplings $y_{3\,31}^e$, $y_{3\,21}^{e}$, $y_{3\,32}^\mu$ and $y_{3\,22}^{\mu}$ are allowed. For simplicity, we will assume all couplings to be real.

Aiming at a minimal working example, we set all the other couplings to zero and consider the following structure for the coupling matrices in our analysis:
%%%%%%%%%%%
\begin{eqnarray}
\begin{aligned}
    & y_1^{\tau} = \left( \begin{array}{ccc} 0 & 0 & 0  \\ 0 & 0 & 0 \\ 0 & 0 & y_{1\, 33}^{\tau} \\ \end{array} \right) \,, \qquad\qquad
    && x_1^{\tau} = \left( \begin{array}{ccc} 0 & 0 & 0 \\ 0 & 0 & x_{1\, 23}^{\tau} \\ 0 & 0 & 0 \\ \end{array} \right) \,, \\
    & y_3^{e} = \left( \begin{array}{ccc} 0 & 0 & 0 \\  y_{3\, 21}^{e} & 0 & 0 \\ y_{3\, 31}^{e} & 0 & 0 \\ \end{array} \right) \,, \qquad\qquad
    && y_3^{\mu} =  \left( \begin{array}{ccc} 0 & 0 & 0 \\ 0 & y_{3\, 22}^{\mu} & 0 \\ 0 & y_{3\, 32}^{\mu} & 0 \\ \end{array} \right) \,.
\end{aligned}\label{eq:S1S3_coulpings}
\end{eqnarray}
%%%%%%%%%%%

The IR FP values for these six non-vanishing couplings are therefore obtained by searching for the simultaneous zeros of their relative beta functions, as given in Sec.~\ref{sec:rge_s1s3}. Only one family of solutions is found to be complying with requirement \emph{i}) listed above, which we report in Table~\ref{tab:s1s3_fp}.
%%%%%%%%%%%
\begin{table}[t!]
    \centering
    \begin{tabular}{cc|ccccc}
        \toprule
        $y_{1\, 33}^{\tau}$ & $x_{1\, 23}^{\tau}$ & $y_{3\, 21}^{e}$ & $y_{3\, 31}^{e}$ & $y_{3\, 22}^{\mu}$ & $y_{3\, 32}^{\mu}$  \\
        \hline
        $0.986$ & $0.871$ & $0.672$ & $0.433$ & $0.672$ & $- 0.433$ \\
        \bottomrule
    \end{tabular}
    \caption{Values for the IR FP of the six non-vanishing LQ couplings defined in Eq.~\eqref{eq:S1S3_coulpings}. Additional solutions obtained via sign permutation are allowed as well, see text for further details. \label{tab:s1s3_fp}}
\end{table}
%%%%%%%%%%%
The solution is characterized by sign ambiguities, meaning that we can simultaneously flip signs of several couplings to find new solutions: for each of the two $S_1^\tau$ couplings both signs are allowed, while for the four $S_3^e$ and $S_3^\mu$ couplings an odd number of them, namely either one or three, has to be negative, with the remaining ones being positive. This means that this family of solution is composed by 32 different scenarios, distinguished by sign permutations.

Unfortunately, this family of solutions is phenomenologically non-viable. On the one hand, the $S_1^\tau$ sector looks promising, with both couplings being $\sim 1$ and hence complying with the size implied by $b\to c$ anomalies and given at Eq.~\eqref{eq:S1_pheno_size}. On the other hand, the $S_3^\ell$ sector has an unacceptable, albeit intriguing, behaviour: indeed, the couplings are connected by the relation $y_{3\, 21}^{e} y_{3\, 31}^{e} = -y_{3\, 22}^{\mu} y_{3\, 32}^{\mu}$, which is in maximal disagreement with requirement \emph{ii}). For this reason, it is not possible to connect the low-energy behaviour of this kind of LQ extension of the SM to the IR FP values of the NP couplings, if trying to address coherently the pattern of deviations in $B$ meson decays.

We therefore move to inspect a more generalized scenario, where six NP fields are allowed in the extension of the SM. In analogy of the three particle scenario, we allow only the following couplings to be non-vanishing:
%%%%%%%%%%%
\begin{eqnarray}\label{eq:S1S3_gen}
\begin{aligned}
    & y_1^{e} = \left( \begin{array}{ccc} 0 & 0 & 0 \\ 0 & 0 & 0 \\ y_{1\, 31}^{e} & 0 & 0 \\ \end{array} \right) \,, \quad
    && y_1^{\mu} = \left( \begin{array}{ccc} 0 & 0 & 0  \\ 0 & 0 & 0 \\ 0 & y_{1\, 32}^{\mu} & 0 \\ \end{array} \right) \,, \quad
    &&& y_1^{\tau} = \left( \begin{array}{ccc} 0 & 0 & 0  \\ 0 & 0 & 0 \\ 0 & 0 & y_{1\, 33}^{\tau} \\ \end{array} \right) \,, \\
    & x_1^{e} = \left( \begin{array}{ccc} 0 & 0 & 0 \\ x_{1\, 21}^{e} & 0 & 0 \\ 0 & 0 & 0 \\ \end{array} \right) \,, \quad
    && x_1^{\mu} = \left( \begin{array}{ccc} 0 & 0 & 0 \\ 0 & x_{1\, 22}^{\mu} & 0 \\ 0 & 0 & 0 \\ \end{array} \right) \,, \quad
    &&& x_1^{\tau} = \left( \begin{array}{ccc} 0 & 0 & 0 \\ 0 & 0 & x_{1\, 23}^{\tau} \\ 0 & 0 & 0 \\ \end{array} \right) \,, \\
    & y_3^{e} = \left( \begin{array}{ccc} 0 & 0 & 0 \\  y_{3\, 21}^{e} & 0 & 0 \\ y_{3\, 31}^{e} & 0 & 0 \\ \end{array} \right) \,, \qquad\qquad
    && y_3^{\mu} =  \left( \begin{array}{ccc} 0 & 0 & 0 \\ 0 & y_{3\, 22}^{\mu} & 0 \\ 0 & y_{3\, 32}^{\mu} & 0 \\ \end{array} \right) \,, \quad
    &&& y_3^{\tau} =  \left( \begin{array}{ccc} 0 & 0 & 0 \\ 0 & 0 & y_{3\, 23}^{\tau} \\ 0 & 0 & y_{3\, 33}^{\tau} \\ \end{array} \right) \,.
\end{aligned}\label{eq:S1S3_gen_coulpings}
\end{eqnarray}
%%%%%%%%%%%

In this new scenario, it is possible to find the following two families of solutions, both complying with requirements \emph{i}) and \emph{ii}). The results are listed in Table~\ref{tab:s1s3_gen_fp}.
%%%%%%%%%%%
\begin{table}[t!]
    \centering
    \scalebox{0.95}{
    \begin{tabular}{cccccc|cccccc}
        \toprule
        $y_{1\, 31}^{e}$ & $x_{1\, 21}^{e}$ & $y_{1\, 32}^{\mu}$ & $x_{1\, 22}^{\mu}$ & $y_{1\, 33}^{\tau}$ & $x_{1\, 23}^{\tau}$ & $y_{3\, 21}^{e}$ & $y_{3\, 31}^{e}$ & $y_{3\, 22}^{\mu}$ & $y_{3\, 32}^{\mu}$ & $y_{3, 23}^{\tau}$ & $y_{3\, 33}^{\tau}$  \\
        \hline
        $0.291$ & $1.006$ & $0.291$ & $1.006$ & $0.291$ & $1.006$ & $0.749$ & $0.172$ & $0.172$ & $0.749$ & $0.664$ & $-0.388$ \\
        $0.291$ & $1.006$ & $0.291$ & $1.006$ & $0.291$ & $1.006$ & $0.172$ & $0.749$ & $0.749$ & $0.171$ & $0.663$ & $-0.388$ \\
        \bottomrule
    \end{tabular}}
    \caption{Values for the IR FP of the twelve non-vanishing LQ couplings defined in Eq.~\eqref{eq:S1S3_gen_coulpings}. Additional solutions obtained via sign permutation are allowed as well, see text for further details. \label{tab:s1s3_gen_fp}}
\end{table}
%%%%%%%%%%%
Similarly to the previous case, also these solutions are characterized by sign ambiguities: concerning the six couplings in the $S_1$ sector, both signs are allowed for each of them, yielding 64 different configurations; on the other hand, concerning the six couplings in the $S_3$ sector, the product of the electron couplings and the one of the muon coupling have to share the same sign, while the product of the tau ones have to be opposite, yielding 16 different configurations. Hence, in total, each family of solutions is composed by 1024 distinct solutions due to sign permutations.

\begin{figure}[t!]
\centering
  \includegraphics[width=0.49\textwidth]{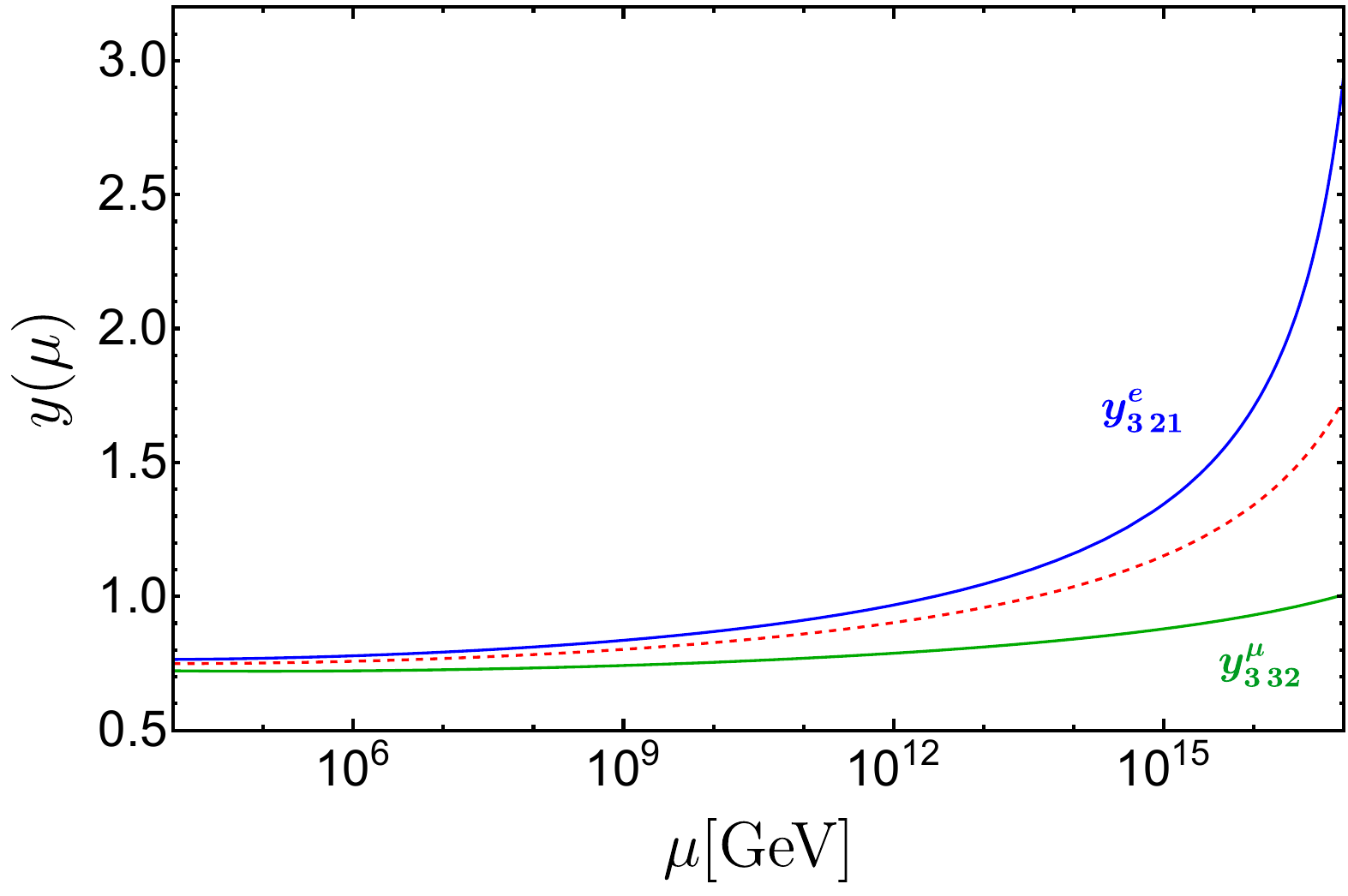}
  \includegraphics[width=0.49\textwidth]{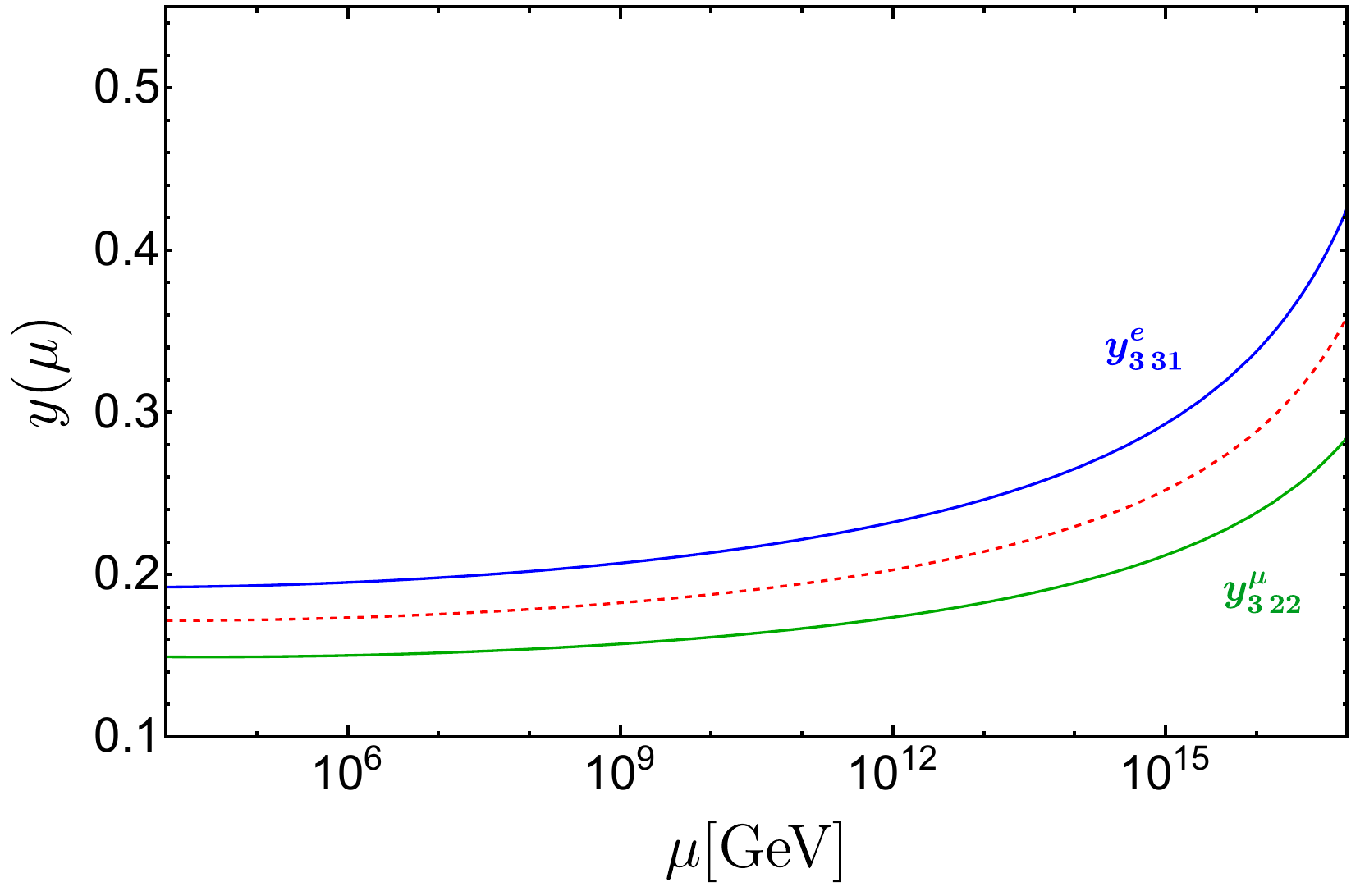}
  \caption{Scenario of Eq.~\eqref{eq:S1S3_gen}: Left panel: running of the couplings $(y_{3\, 21}^{e}$ and $y_{3\, 32}^{\mu})$ from the high-scale to the low-scale; the FP solution is given in dashed red. Right panel: running of the couplings $(y_{3\, 31}^{e}$ and $y_{3\, 22}^{\mu})$ from the high-scale to the low-scale; the FP solution is given in dashed red.\label{fig:S1S3_FP}
  }
\end{figure}

It is interesting to highlight that requirement \emph{ii}) is not accidentally fulfilled, but it is met due to the pairs of couplings $(y_{3\, 21}^{e},y_{3\, 32}^{\mu})$ and $(y_{3\, 31}^{e}, y_{3\, 22}^{\mu})$ sharing the same IR FP, respectively. Hence, the low-energy LFU observed in $b\to s \ell\ell$ transitions can be elegantly explained due to a dynamical behaviour, with the couplings not having to share the same pattern at the high-scale. An example of this behaviour can be seen in the two panels of Fig.~\ref{fig:S1S3_FP}, where the four couplings are taken to be different at the high-scale but are attracted to the same two low-scale values, which corresponds to the FP solution of their beta functions. In particular, the four couplings are evolved from $\mu_h=10^{17}$ GeV down to $\mu_l=10^4$ GeV, and their values are reported below:
\begin{eqnarray}
\begin{aligned}
    y_{3\, 21}^{e}(\mu_h) &= 2.95 \,, \qquad
    y_{3\, 32}^{\mu}(\mu_h) = 1.00 \,, \qquad
    y_{3\, 31}^{e}(\mu_h) = 0.426 \,, \qquad
    y_{3\, 22}^{\mu}(\mu_h) = 0.284 \,, \\
    y_{3\, 21}^{e}(\mu_l) &= 0.765 \,, \quad\,\,\,
    y_{3\, 32}^{\mu}(\mu_l) = 0.722 \,, \qquad
    y_{3\, 31}^{e}(\mu_l) = 0.192 \,, \qquad
    y_{3\, 22}^{\mu}(\mu_l) = 0.149 \,.
\end{aligned}
\end{eqnarray}
Note that the beta functions also depend on SM couplings which depend on the renormalization scale $\mu$. Therefore the FP solution of the LQ couplings is not a constant line.

%%%%%%%%%%%
%%%%%%%%%%%
Employing now the FP values for the LQ couplings and inverting Eq.~\eqref{eq:S3_CL}, one infers the scale for the LQ masses to be 
%%%%%%%%%%%%
\begin{equation}
M_{S_3^e} = M_{S_3^\mu} = 14.5 \sqrt{\frac{0.04}{|V_{tb} V_{ts}^*|}}\sqrt{\frac{-0.4}{C_L^U}}\, \text{TeV}\, . \label{lqm}
\end{equation}
%%%%%%%%%%%%%
It is worthwhile to compare this with the prediction from the UV FP analysis in Ref.~\cite{Kowalska:2020gie}. As a first remark, no choice of UV boundary conditions can lead to IR values of the LQ couplings with a positive beta function, so that at least one of the two couplings entering the $b\to s \ell^+\ell^-$ amplitude must be (in magnitude) below its IR fixed point. Indeed, the product of the low-energy values of these couplings found in Ref.~\cite{Kowalska:2020gie} is smaller than ours, with the larger coupling similar in size to ours. Thus also the corresponding value for $M_{S_3^{e,\mu}}$ inferred from \eq{lqm} is smaller. \eq{lqm} sets an upper bound on $M_{S_3^{e,\mu}}$ for any choice of UV boundary condition at the GUT or Planck scale. At the same time, any $S_3^{e,\mu}$ discovery with a mass close to the value in \eq{lqm} will imply LQ couplings close to their IR fixed point, so that their high-energy values are unpredictable. The lower bound on the products of the couplings and thereby on $M_{S_3^{e,\mu}}$ found in Ref.~\cite{Kowalska:2020gie} is much more sensitive to the UV boundary condition, as it results from UV values of couplings too small to ever reach the IR fixed point. In other words: the smaller $M_{S_3^{e,\mu}}$, the more information can be inferred on the UV values of the couplings from low-energy measurements. 

It is also interesting to notice that, with these values for the LQ couplings and mass, the expected impact to the $b\to s\nu\bar\nu$ transitions ratio $R_{K^{(*)}}^{\nu\bar\nu}$ defined at Eq.~\eqref{eq:RKnunu} reads $R_{K^{(*)}}^{\nu\bar\nu} \simeq 1.1$, hence not large enough to meaningfully also reduce the recent Belle II excess for $R_{K}^{\nu\bar\nu}$~\cite{BelleIISemKnunu}.

Finally, the emergence of an electron-muon universality implies also a strong and precise prediction for the tau couplings, whose product is characterized by an opposite sign w.r.t. the light leptons. In particular, both FP solutions predict an enhancement in the tau sector (opposite to the suppression implied by present $b\to s$ data in light leptons) corresponding to $C_L^{\tau}(\mu_b) \sim 0.8$, if one assumes $M_{S_3^e} = M_{S_3^\mu} = M_{S_3^\tau}$.

\begin{figure}[t!]
\centering
  \includegraphics[width=0.45\textwidth]{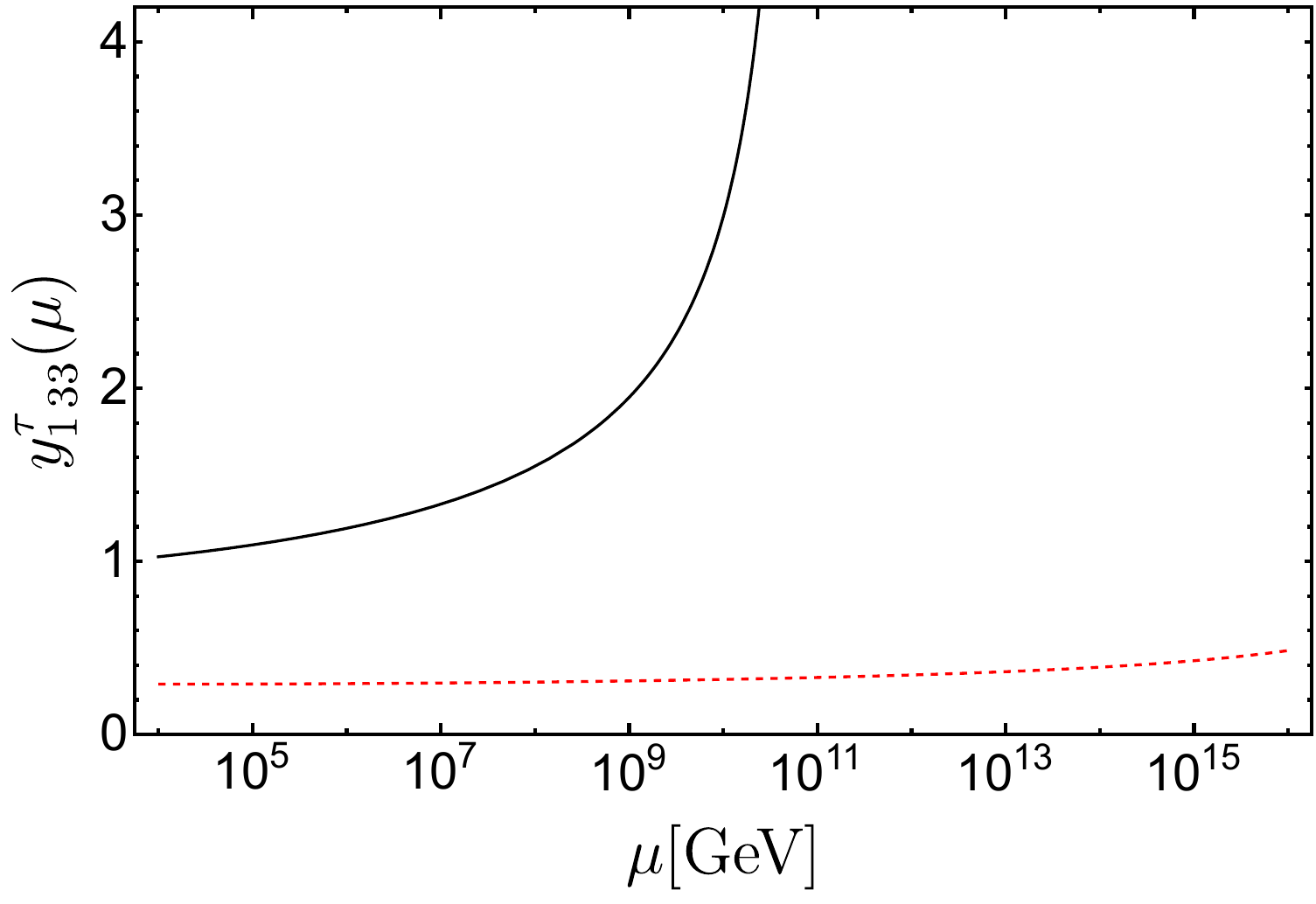}
  \caption{Emergence of a Landau pole in the running of the coupling $y_{1\, 23}^{\tau}$}, when a value compatible with $b\to c\tau\bar\nu$ data is assumed at the low-scale. The FP solution is given in dashed red. \label{fig:S1S3_LP}
\end{figure}

The situation is different in the $b\to c$ sector: indeed, the FP solution for the $S_1^\tau$ coupling yield $y_{1\, 33}^{\tau} x_{1\, 23}^{\tau} \simeq -0.3$, where the freedom on the coupling signs allows us to choose $x_{1\, 23}^{\tau} \simeq -1$; however, when confronting this value with  Eq.~\eqref{eq:S1_pheno_size}, in order to address the anomalies in $b\to c$ transitions $S_1^\tau$ would be required to have a mass equal to $M_{S_1^\tau}\sim 0.8\,\text{TeV}$, value which has already been excluded by direct searches at LHC.\footnote{The NP contribution to $C_{V_L}^{e,\mu}$ coming from non-vanishing couplings of $S_1^e$ and $S_1^\mu$ are strongly constrained, see e.g. Ref.~\cite{Fedele:2022iib} and references therein. In order to suppress such undesirable effects, the masses of these two LQs are considered to be sensitively heavier than the scale of $M_{S_1^\tau}$.} This implies that, if one would like to address the current experimental situation in this sector as well, the value for $y_{1\, 23}^{\tau}$ at the low scale $\mu_l=10^4$ GeV has to be taken well above the FP solution, namely equal to $y_{1\, 23}^{\tau}\sim 1$. In turn, this implies the emergence of a Landau pole at a scale around $\mu \sim 10^{11}\,\text{GeV}$, which is the scale where $y_{1\, 23}^{\tau}$ diverges as can be observed in Fig.~\ref{fig:S1S3_LP}.

To conclude we have obtained that, when extending the SM sector with the 6 scalar LQs $S_1^e$, $S_1^\mu$, $S_1^\tau$, $S_3^e$, $S_3^\mu$ and $S_3^\tau$, thanks to the IR FP behaviours of their couplings it is possible to explain the observed pattern of anomalies in $b \to s\ell\ell$ transitions by introducing $S_3^e$ and $S_3^\mu$ LQs with masses at the $\sim 10\,\text{TeV}$ scale and arbitrary high-scale couplings; on the other hand, in order to address the experimental picture in $b\to c\tau\bar\nu$ transitions as well, a value above the FP solution is required for one of the couplings, inducing an upper limit to the LQU scale equal to $M_{\rm LQU}\lesssim 10^{11}\,\text{GeV}$, which is far below the GUT scale and corroborates ideas of multi-step unification~\cite{FileviezPerez:2013zmv}.

%%%%%%%%%%%
%%%%%%%%%%%

%%%%%%%%%%%%%%%%%%%%%%%%%%%%%%%%%

\subsection{The \texorpdfstring{$\mathbf{(R_2,S_3)}$}{(R2,S3)} Extension}\label{sec:fp_r2s3}
We move on to the study of the SM extended by one doublet LQ, $R_2^\tau$, and two copies of the triplet one, $S_3^e$ and $S_3^\mu$. Once again, the triplet LQs are employed to obtain the desired low-energy effect in $b\to s\ell^+\ell^-$ transitions by means of non-vanishing values for the couplings $y_{3\,31}^e$, $y_{3\,21}^{e}$, $y_{3\,32}^\mu$ and $y_{3\,22}^{\mu}$. On the other hand, following now Sec.~\ref{sec:R2intro}, we adopt the doublet LQ in order to explain the $b\to c\tau\bar\nu$ decays phenomenology, which require the presence of the $y_{2\,23}^\tau$ and $x_{2\,33}^{\tau}$, with their product being imaginary as detailed in requirement \emph{iii}). We therefore allow the two $R_2^\tau$ couplings to be complex.

The minimal set of non-vanishing couplings required by our analysis is therefore:
%%%%%%%%%%%
\begin{eqnarray}
\begin{aligned}
    & y_2^{\tau} = \left( \begin{array}{ccc} 0 & 0 & 0  \\ 0 & 0 & y_{2\, 23}^{\tau} \\ 0 & 0 & 0 \\ \end{array} \right) \,, \qquad\qquad
    && x_2^{\tau} = \left( \begin{array}{ccc} 0 & 0 & 0 \\ 0 & 0 & 0 \\ 0 & 0 & x_{2\, 33}^{\tau} \\ \end{array} \right) \,, \\
    & y_3^{e} = \left( \begin{array}{ccc} 0 & 0 & 0 \\  y_{3\, 21}^{e} & 0 & 0 \\ y_{3\, 31}^{e} & 0 & 0 \\ \end{array} \right) \,, \qquad\qquad
    && y_3^{\mu} =  \left( \begin{array}{ccc} 0 & 0 & 0 \\ 0 & y_{3\, 22}^{\mu} & 0 \\ 0 & y_{3\, 32}^{\mu} & 0 \\ \end{array} \right) \,.
\end{aligned}\label{eq:R2S3_coulpings}
\end{eqnarray}
%%%%%%%%%%% 

In a similar fashion to the previous scenario, we look now for the simultaneous zeros of the couplings beta functions, as given in Sec.~\ref{sec:rge_r2s3}. In this case, two families of solutions are found to be complying with requirements \emph{i}) and \emph{iii}) listed above, identified by which of the two $R_2^\tau$ couplings is purely imaginary, and listed in Table~\ref{tab:r2s3_fp}.
%%%%%%%%%%%
\begin{table}[t!]
    \centering
    \begin{tabular}{cc|ccccc}
        \toprule
        $y_{2\, 23}^{\tau}$ & $x_{2\, 33}^{\tau}$ & $y_{3\, 21}^{e}$ & $y_{3\, 31}^{e}$ & $y_{3\, 22}^{\mu}$ & $y_{3\, 32}^{\mu}$  \\
        \hline
        $1.094\, i$ & $0.783$ & $0.654$ & $0.472$ & $0.654$ & $-0.472$ \\
        $1.094$ & $0.783\, i$ & $0.654$ & $0.472$ & $0.654$ & $-0.472$ \\
        \bottomrule
    \end{tabular}
    \caption{Values for the IR FP of the six non-vanishing LQ couplings defined in Eq.~\eqref{eq:R2S3_coulpings}. Additional solutions obtained via sign permutation are allowed as well, see text for further details. \label{tab:r2s3_fp}}
\end{table}
%%%%%%%%%%%
These solutions are both characterized by the same sign ambiguities: for each of the two $R_2^\tau$ couplings both signs are allowed, while for the four $S_3^e$ and $S_3^\mu$ couplings an odd number of them, namely either one or three, has to be negative, with the remaining ones being positive. This means that both families of solution are composed by 32 different scenarios each, distinguished by sign permutations.

The minimal scenario is not found to be phenomenologically viable in this configuration as well. A maximal disagreement with requirement \emph{ii}) is again present, with $y_{3\, 21}^{e} y_{3\, 31}^{e} = -y_{3\, 22}^{\mu} y_{3\, 32}^{\mu}$, invalidating an explanation to $b\to s\ell^+\ell^-$ data. Moreover, even if requirement \emph{iii}) is fulfilled, the FP values for the $R_2^\tau$ couplings are not acceptable if one wants to address anomalies in $b \to c\tau\bar\nu$ transitions: indeed, the product of the two couplings is well below $\sim 2$ (in modulus), which is the value required to have a mass for the LQ not excluded by direct searches, see Sec.~\ref{sec:R2intro}.

We therefore move to inspect a more generalized scenario, where six NP fields are allowed in the extension of the SM. In analogy of the three particle scenario, we allow only the following couplings to be non-vanishing:
%%%%%%%%%%%
\begin{eqnarray}
\begin{aligned}
    & y_2^{e} = \left( \begin{array}{ccc} 0 & 0 & 0  \\ y_{2\, 21}^{e} & 0 & 0 \\ 0 & 0 & 0 \\ \end{array} \right) \,, \quad
    && y_2^{\mu} = \left( \begin{array}{ccc} 0 & 0 & 0  \\ 0 & y_{2\, 22}^{\mu} & 0 \\ 0 & 0 & 0 \\ \end{array} \right) \,, \quad
    &&& y_2^{\tau} = \left( \begin{array}{ccc} 0 & 0 & 0  \\ 0 & 0 & y_{2\, 23}^{\tau} \\ 0 & 0 & 0 \\ \end{array} \right) \,, \\
    & x_2^{e} = \left( \begin{array}{ccc} 0 & 0 & x_{2\, 13}^{e} \\ 0 & 0 & 0 \\ 0 & 0 & 0 \\ \end{array} \right) \,, \quad
    && x_2^{\mu} = \left( \begin{array}{ccc} 0 & 0 & 0 \\ 0 & 0 & x_{2\, 23}^{\mu} \\ 0 & 0 & 0 \\ \end{array} \right) \,, \quad
    &&& x_2^{\tau} = \left( \begin{array}{ccc} 0 & 0 & 0 \\ 0 & 0 & 0 \\ 0 & 0 & x_{2\, 33}^{\tau} \\ \end{array} \right) \,, \\
    & y_3^{e} = \left( \begin{array}{ccc} 0 & 0 & 0 \\  y_{3\, 21}^{e} & 0 & 0 \\ y_{3\, 31}^{e} & 0 & 0 \\ \end{array} \right) \,, \qquad\qquad
    && y_3^{\mu} =  \left( \begin{array}{ccc} 0 & 0 & 0 \\ 0 & y_{3\, 22}^{\mu} & 0 \\ 0 & y_{3\, 32}^{\mu} & 0 \\ \end{array} \right) \,, \quad
    &&& y_3^{\tau} =  \left( \begin{array}{ccc} 0 & 0 & 0 \\ 0 & 0 & y_{3\, 23}^{\tau} \\ 0 & 0 & y_{3\, 33}^{\tau} \\ \end{array} \right) \,.
\end{aligned}\label{eq:R2S3_gen_coulpings}
\end{eqnarray}
%%%%%%%%%%%

We find also in this generalized scenario two families of solution complying with requirements \emph{i}) and \emph{iii}), according to which is the $R_2^\tau$ coupling to assume imaginary values. The results are reported in Table~\ref{tab:r2s3_gen_fp}.
%%%%%%%%%%%
\begin{table}[t!]
    \centering
    \scalebox{0.9}{
    \begin{tabular}{cccccc|cccccc}
        \toprule
        $y_{2\, 21}^{e}$ & $x_{2\, 13}^{e}$ & $y_{2\, 22}^{\mu}$ & $x_{2\, 23}^{\mu}$ & $y_{2\, 23}^{\tau}$ & $x_{2\, 33}^{\tau}$ & $y_{3\, 21}^{e}$ & $y_{3\, 31}^{e}$ & $y_{3\, 22}^{\mu}$ & $y_{3\, 32}^{\mu}$ & $y_{3, 23}^{\tau}$ & $y_{3\, 33}^{\tau}$   \\
        \hline
        $0.584$ & $0.837$ & $0.584$ & $0.837$ & $0.584\, i$ & $0.837$ & $0.679$ & $0.181$ & $0.679$ & $0.181$ & $0.521$ & $-0.472$ \\
        $0.584$ & $0.837$ & $0.584$ & $0.837$ & $0.584$ & $0.837\, i$ & $0.679$ & $0.181$ & $0.679$ & $0.181$ & $0.521$ & $-0.472$ \\
        \bottomrule
    \end{tabular}}
    \caption{Values for the IR FP of the twelve non-vanishing LQ couplings defined in Eq.~\eqref{eq:R2S3_gen_coulpings}. Additional solutions obtained via sign permutation are allowed as well, see text for further details. \label{tab:r2s3_gen_fp}}
\end{table}
%%%%%%%%%%%
In a similar fashion to what observed in the previous Section, these solutions are characterized by the same sign ambiguities: both signs are allowed for each of the $R_2$ couplings, yielding 64 different configurations, while the sign has to be the same for the product of both $S_3^e$ and $S_3^\mu$ couplings, respectively, and opposite for the product of $S_3^\tau$ ones, yielding 16 different configurations. Summarizing, each family of solutions is composed by 1024 distinct solutions due to sign permutations.

These found solutions share a similar phenomenology to the ones found in the general case studied in Sec.~\ref{sec:fp_s1s3}. Indeed, requirement \emph{ii}) is fulfilled since the pairs of couplings $(y_{3\, 21}^{e},y_{3\, 32}^{\mu})$ and $(y_{3\, 31}^{e}, y_{3\, 22}^{\mu})$ share the same IR FP, respectively: we therefore obtain that, also in this scenario, the low-energy LFU behaviour required to address $b\to s\ell^+\ell^-$ data can be ascribed to a FP origin. Moreover, the inferred value from Eq.~\eqref{eq:S3_CL} for the LQ masses now reads 
%%%%%%%%%%%%
\begin{equation}
M_{S_3^e} = M_{S_3^\mu} = 14.1 \sqrt{\frac{0.04}{|V_{tb} V_{ts}^*|}}\sqrt{\frac{-0.4}{C_L^U}}\, \text{TeV}\,,
\end{equation}
%%%%%%%%%%%%%
Finally, the prediction for $R_{K^{(*)}}^{\nu\bar\nu}$ defined at Eq.~\eqref{eq:RKnunu} is equal to $R_{K^{(*)}}^{\nu\bar\nu} \simeq 1.1$, again not large enough to significantly reduce the Belle II excess in $R_{K}^{\nu\bar\nu}$~\cite{BelleIISemKnunu}. To conclude, in this scenario as well we observe an opposite behaviour in the tau sector compared to the one observed for the light leptons, with the product of the tau couplings showing again an opposite sign and a prediction for the NP effect in this sector equal to $C_L^{\tau}(\mu_b) \sim 0.8$, again in the case of degenerate masses.

On the other hand, the situation in the $b\to c$ sector is again different: the tau couplings product reads here in both cases $|y_{1\, 33}^{\tau} x_{1\, 23}^{\tau}| \simeq 0.5$, again too small to reproduce the desired phenomenology. It is nevertheless interesting to investigate, in this scenario as well, the implications of taking values for the couplings above the FP solution. Indeed, taking for both couplings a value in modulus of the order $\sim\sqrt2$, we can observe again the emergence of a Landau pole, found this time at the scale $\mu \sim 10^{8}\,\text{GeV}$.

In conclusion, we observed that the study of IR FPs yield interesting phenomenological implications when the SM is extended with the 6 scalar LQs $R_2^e$, $R_2^\mu$, $R_3^\tau$, $S_3^e$, $S_3^\mu$ and $S_3^\tau$. In particular, the LFU required at the low scale for $b\to s\ell^+\ell^-$ transitions can be obtained by the FP behaviours of the $S_3^e$ and $S_3^\mu$ couplings, whose masses are require to be at the $\sim 10\,\text{TeV}$ scale while no additional conditions are requested for the coupling values at the high-scale. On the other hand, a value for the $R_3^\tau$ couplings is required to be above the FP solution when confronting with $b\to c\tau\bar\nu$, implying the emergence of an upper limit to the LQU scale equal to $M_{\rm LQU}\lesssim 10^{8}\,\text{GeV}$.

%%%%%%%%%%%%%%%%%%%%%%%%%%%%%%%%%

\subsection{The \texorpdfstring{$\mathbf{S_3}$}{S3} Extension}\label{sec:fp_s3}
We conclude this Section by studying the FP solution of SM extensions of $S_3$ LQs only, even if in this scenario $b\to c\tau\bar\nu$ data cannot be explained. Indeed, motivated by the findings of Secs.~\ref{sec:fp_s1s3} and~\ref{sec:fp_r2s3}, it is interesting to investigate whether the dynamical emergence of LFU in the $S_3$ contributions to $b\to s\ell^+\ell^-$ transitions arises only in the presence of additional LQs in the theory as well, or it is an exclusive feature of the triplet LQs. 

Following the approach of the previous analyses, we start our study from the scenario where only $S_3^e$ and $S_3^\mu$ are added to the theory, with non-vanishing values for the couplings $y_{3\,31}^e$, $y_{3\,21}^{e}$, $y_{3\,32}^\mu$ and $y_{3\,22}^{\mu}$. The only found family of solutions complying with requirement \emph{i}) is given in Table~\ref{tab:s3_fp}.
%%%%%%%%%%%
\begin{table}[t!]
    \centering
    \begin{tabular}{ccccc}
        \toprule
        $y_{3\, 21}^{e}$ & $y_{3\, 31}^{e}$ & $y_{3\, 22}^{\mu}$ & $y_{3\, 32}^{\mu}$  \\
        \hline
        $0.622$ & $0.533$ & $0.622$ & $-0.533$ \\
        \bottomrule
    \end{tabular}
    \caption{Values for the IR FP of the four non-vanishing LQ couplings entering in $y_3^e$ and $y_3^\mu$ matrices defined in Eq.~\eqref{eq:S1S3_coulpings}. Additional solutions obtained via sign permutation are allowed as well, see text for further details. \label{tab:s3_fp}}
\end{table}
%%%%%%%%%%%
%%%%%%%%%%%
\begin{table}[b!]
    \centering
    \begin{tabular}{cccccc}
        \toprule
        $y_{3\, 21}^{e}$ & $y_{3\, 31}^{e}$ & $y_{3\, 22}^{\mu}$ & $y_{3\, 32}^{\mu}$ & $y_{3\, 23}^{\tau}$ & $y_{3\, 33}^{\tau}$  \\
        \hline
        $0.760$ & $0.189$ & $0.191$ & $0.759$ & $0.639$ & $-0.452$ \\
        $0.189$ & $0.760$ & $0.759$ & $0.191$ & $0.639$ & $-0.452$ \\
        \bottomrule
    \end{tabular}
    \caption{Values for the IR FP of the six non-vanishing LQ couplings entering in $y_3^e$, $y_3^\mu$ and $y_3^\tau$ matrices defined in Eq.~\eqref{eq:S1S3_gen_coulpings}. Additional solutions obtained via sign permutation are allowed as well, see text for further details. \label{tab:s3_gen_fp}}
\end{table}
%%%%%%%%%%%
These solutions share the same features of the scenarios studied in Secs.~\ref{sec:fp_s1s3} and~\ref{sec:fp_r2s3}, when only two copies of the triplet LQ were allowed. In particular, once again an odd number of minus signs is allowed for the four couplings, yielding to 8 different solutions distinguished by sign permutations. Moreover, requirement \emph{ii}) is again not fulfilled, due to the emerging feature $y_{3\, 21}^{e} y_{3\, 31}^{e} = -y_{3\, 22}^{\mu} y_{3\, 32}^{\mu}$.

Moving on to the generalized case where the $S_3^\tau$ LQ is included as well, the solutions following requirements \emph{i}) and \emph{ii}) are reported in Table~\ref{tab:s3_gen_fp}. Also these solutions are qualitatively similar to the ones studied in the previous sections, when all three copies of $S_3$ were allowed. Indeed, each family of solutions is characterized by the same sign ambiguity, with the sign for both $S_3^e$ and $S_3^\mu$ couplings products having to be the same, respectively, and opposite for $S_3^\tau$ one. Moreover, requirement \emph{ii}) is again dynamically fulfilled and the following LQ masses are predicted:
%%%%%%%%%%%%
\begin{equation}
M_{S_3^e} = M_{S_3^\mu} = 15.3 \sqrt{\frac{0.04}{|V_{tb} V_{ts}^*|}}\sqrt{\frac{-0.4}{C_L^U}}\, \text{TeV}\,,
\end{equation}
%%%%%%%%%%%%%
We therefore find that, in order to obtain this feature, the additional presence of singlet or doublet LQs in the theory is not required. It is worth to mention that, if one would employ a different version of requirement \emph{ii}) requesting, e.g., universality among electrons and taus, those two sectors would be the ones having couplings with the same product, with the product of muon ones being different and opposite in sign.

%%%%%%%%%%%%%%%%%%%%%%%%%%%%%%%%%

\section{Conclusions}\label{sec:summary}

In this paper, we studied the implications of RGE effects to LQ couplings to fermions in selected BSM scenarios. A popular way to address the recent discrepancies observed in several observables in the decays $b\to c\tau \bar\nu$ and $b\to s \ell^+\ell^-$ with $\ell=e,\mu$ consists of extending the SM sectors by means of scalar LQs. In particular, the minimal subset of required new fields includes the presence of triplet LQs $S_3^e$ and $S_3^\mu$, coupled with equal strength to electrons and muons, respectively, and of either the singlet LQ $S_1^\tau$ or the doublet LQ $R_2^\tau$ coupled to taus. Indeed, the former pair of LQs are required to explain anomalies in $b\to s \mu^+\mu^-$ without violating the reanalysed results of the LFUV ratios $R_{K^{(*)}}\sim 1$, while the latter LQ is necessary to address anomalies in the $b\to c\tau \bar\nu$ sector.

While these new fields are expected to live at scales between a few and a few tens of TeV, one expects a large mass gap between the scales $M_{\rm LQ}$ and $M_{\rm QLU}$, the scale where the LQs are generated within a theory of quark-lepton unification, because gauge bosons coupling a quark to a lepton must be very heavy. The presence of this large scale separation therefore implies the possibility that the pattern of values of the LQ Yukawa couplings observed at the $B$ meson decay scale (when employing this kind of SM extensions to address the anomalous data) has a dynamical origin stemming from the infrared behaviour of the RG evolution rather than from symmetry properties. In particular, the possibility of such an explanation of the LFU pattern inferred for the $S_3^e$ and $S_3^\mu$ couplings from $R_{K^{(*)}}\sim 1$ is tantalizing. To this end, we studied the IR fixed-point (FP) solutions of the beta functions of the LQ couplings and inspected their phenomenology using low-energy flavour data.

We found interesting phenomenological solutions in several scenarios. In particular, every time that the SM is extended by three triplet LQs coupled each to a specific lepton, namely $S_3^e$, $S_3^\mu$ and $S_3^\tau$, we find IR fixed point solutions for which the product of $S_3^e$ couplings is equal to the one of the $S_3^\mu$ couplings, so that electron-muon universality can arise dynamically. Such universality is therefore independent from the values assumed by the couplings at the high scale, as shown in an illustrative example in Fig.~\ref{fig:S1S3_FP}, and occurs both when the triplet LQs are the only fields added to the SM, or come together with either three singlet LQs or three doublet ones, namely $S_1^e$, $S_1^\mu$ and $S_1^\tau$, or $R_2^e$, $R_2^\mu$ and $R_2^\tau$, respectively. Moreover, a prediction for the masses of $S_3^{e,\mu}$ between 14 and 15 TeV is obtained (according to the specific scenario), together with a 10\% enhancement in $R_{K^{(*)}}^{\nu\bar\nu}$. While LQs with these masses are beyond the reach of current collider searches, such an increase in $R_{K^{(*)}}^{\nu\bar\nu}$ is within the reach of the Belle II experiment. Furthermore, electron-muon universality implies an IR FP for the product of $S_3^\tau$ couplings with opposite sign, enhancing $C_L^\tau$ over $C_L^{e,\mu}$.

A widely studied research line of flavour physics aims at an understanding of quark masses and Yukawa couplings in terms of broken flavour symmetries. Once LQs are included in such a theory of flavour, it is mandatory to address their representations w.r.t.\ the chosen flavour group (see Ref.~\cite{Varzielas:2015iva} for a comprehensive study). The experimental result $R_{K^{(*)}}\sim 1$ teaches us that  LQs addressing $b\to s\ell^+\ell^-$ anomalies must come in multiple copies distinguished by the lepton flavour number to avoid dangerously large contributions to $\mu\to e$ transitions. Their mass matrix (for the first two generations) must be close to the unit matrix to both avoid  $\mu\to e$ transitions emerging from rotations between weak and mass eigenstates and to accommodate $R_{K^{(*)}}\sim 1$ without the tuning of masses against couplings. This suggest that the LQ mass matrix obeys the SU(2) flavour symmetry of the lepton sector exactly. The results of our paper imply that no symmetry model building for the LQ couplings is necessary: if they are close to their IR fixed-points, LFU emerges dynamically. In this case it will be extremely difficult to gain any insight into their high-energy values, because even small uncertainties of the low-energy values inflate to large errors at high scales (see \fig{fig:S1S3_FP}) making predictive flavour model-building for these couplings impossible.

On the contrary, a dynamical origin has not been found for the couplings required to address anomalies in $b\to c\tau \bar\nu$ decays. Indeed, both in the singlet and in the doublet scenario the IR FP for the relevant couplings have been always found to be below the implied values from low-energy data. Nevertheless, such findings are of phenomenological interest as well, since couplings values above the IR FP imply the emergence of a Landau pole at the scale $\mu \sim 10^{11}\,\text{GeV}$ or $\mu \sim 10^{8}\,\text{GeV}$, depending on whether the SM is extended by scalar or doublet LQs, respectively. This scale can therefore be interpreted as an upper limit on $M_{\rm QLU}$, giving an upper bound on the energy scale where quark-lepton unification should occour.

\acknowledgments  This research was supported by
Deutsche Forschungsgemeinschaft (DFG, German Research
Foundation)  within the Collaborative Research Center
\emph{Particle Physics Phenomenology after the Higgs Discovery (P3H)}\ 
(project no.~396021762 – TRR 257).

\bibliographystyle{JHEP}
\bibliography{hepbiblio}

\providecommand{\href}[2]{#2}\begingroup\raggedright\begin{thebibliography}{10}

\bibitem{LHCb:2014cxe}
{\scshape LHCb} collaboration, \emph{{Differential branching fractions and
  isospin asymmetries of $B \to K^{(*)} \mu^+ \mu^-$ decays}},
  \href{https://doi.org/10.1007/JHEP06(2014)133}{\emph{JHEP} {\bfseries 06}
  (2014) 133} [\href{https://arxiv.org/abs/1403.8044}{{\ttfamily 1403.8044}}].

\bibitem{LHCb:2015wdu}
{\scshape LHCb} collaboration, \emph{{Angular analysis and differential
  branching fraction of the decay $B^0_s\to\phi\mu^+\mu^-$}},
  \href{https://doi.org/10.1007/JHEP09(2015)179}{\emph{JHEP} {\bfseries 09}
  (2015) 179} [\href{https://arxiv.org/abs/1506.08777}{{\ttfamily
  1506.08777}}].

\bibitem{LHCb:2021zwz}
{\scshape LHCb} collaboration, \emph{{Branching Fraction Measurements of the
  Rare $B^0_s\rightarrow\phi\mu^+\mu^-$ and $B^0_s\rightarrow
  f_2^\prime(1525)\mu^+\mu^-$- Decays}},
  \href{https://doi.org/10.1103/PhysRevLett.127.151801}{\emph{Phys. Rev. Lett.}
  {\bfseries 127} (2021) 151801}
  [\href{https://arxiv.org/abs/2105.14007}{{\ttfamily 2105.14007}}].

\bibitem{Khodjamirian:2010vf}
A.~Khodjamirian, T.~Mannel, A.~Pivovarov and Y.-M.~Wang, \emph{{Charm-loop
  effect in $B \to K^{(*)} \ell^{+} \ell^{-}$ and $B\to K^*\gamma$}},
  \href{https://doi.org/10.1007/JHEP09(2010)089}{\emph{JHEP} {\bfseries 09}
  (2010) 089} [\href{https://arxiv.org/abs/1006.4945}{{\ttfamily 1006.4945}}].

\bibitem{Khodjamirian:2012rm}
A.~Khodjamirian, T.~Mannel and Y.M.~Wang, \emph{{$B \to K \ell^{+}\ell^{-}$
  decay at large hadronic recoil}},
  \href{https://doi.org/10.1007/JHEP02(2013)010}{\emph{JHEP} {\bfseries 02}
  (2013) 010} [\href{https://arxiv.org/abs/1211.0234}{{\ttfamily 1211.0234}}].

\bibitem{LHCb:2013ghj}
{\scshape LHCb} collaboration, \emph{{Measurement of Form-Factor-Independent
  Observables in the Decay $B^{0} \to K^{*0} \mu^+ \mu^-$}},
  \href{https://doi.org/10.1103/PhysRevLett.111.191801}{\emph{Phys. Rev. Lett.}
  {\bfseries 111} (2013) 191801}
  [\href{https://arxiv.org/abs/1308.1707}{{\ttfamily 1308.1707}}].

\bibitem{LHCb:2015svh}
{\scshape LHCb} collaboration, \emph{{Angular analysis of the $B^{0} \to K^{*0}
  \mu^{+} \mu^{-}$ decay using 3 fb$^{-1}$ of integrated luminosity}},
  \href{https://doi.org/10.1007/JHEP02(2016)104}{\emph{JHEP} {\bfseries 02}
  (2016) 104} [\href{https://arxiv.org/abs/1512.04442}{{\ttfamily
  1512.04442}}].

\bibitem{LHCb:2020lmf}
{\scshape LHCb} collaboration, \emph{{Measurement of $CP$-Averaged Observables
  in the $B^{0}\rightarrow K^{*0}\mu^{+}\mu^{-}$ Decay}},
  \href{https://doi.org/10.1103/PhysRevLett.125.011802}{\emph{Phys. Rev. Lett.}
  {\bfseries 125} (2020) 011802}
  [\href{https://arxiv.org/abs/2003.04831}{{\ttfamily 2003.04831}}].

\bibitem{LHCb:2020gog}
{\scshape LHCb} collaboration, \emph{{Angular Analysis of the $B^{+}\rightarrow
  K^{\ast+}\mu^{+}\mu^{-}$ Decay}},
  \href{https://doi.org/10.1103/PhysRevLett.126.161802}{\emph{Phys. Rev. Lett.}
  {\bfseries 126} (2021) 161802}
  [\href{https://arxiv.org/abs/2012.13241}{{\ttfamily 2012.13241}}].

\bibitem{hep-ph/0310219}
G.~Hiller and F.~Kruger, \emph{{More model independent analysis of $b \to s$
  processes}}, \href{https://doi.org/10.1103/PhysRevD.69.074020}{\emph{Phys.
  Rev.} {\bfseries D69} (2004) 074020}
  [\href{https://arxiv.org/abs/hep-ph/0310219}{{\ttfamily hep-ph/0310219}}].

\bibitem{LHCb:2022qnv}
{\scshape LHCb} collaboration, \emph{{Test of lepton universality in $b
  \rightarrow s \ell^+ \ell^-$ decays}},
  \href{https://arxiv.org/abs/2212.09152}{{\ttfamily 2212.09152}}.

\bibitem{LHCb:2022zom}
{\scshape LHCb} collaboration, \emph{{Measurement of lepton universality
  parameters in $B^+\to K^+\ell^+\ell^-$ and $B^0\to K^{*0}\ell^+\ell^-$
  decays}},  \href{https://arxiv.org/abs/2212.09153}{{\ttfamily 2212.09153}}.

\bibitem{BaBar:2012obs}
{\scshape BaBar} collaboration, \emph{{Evidence for an excess of $\bar{B} \to
  D^{(*)} \tau^-\bar{\nu}_\tau$ decays}},
  \href{https://doi.org/10.1103/PhysRevLett.109.101802}{\emph{Phys. Rev. Lett.}
  {\bfseries 109} (2012) 101802}
  [\href{https://arxiv.org/abs/1205.5442}{{\ttfamily 1205.5442}}].

\bibitem{Belle:2019rba}
{\scshape Belle} collaboration, \emph{{Measurement of $\mathcal{R}(D)$ and
  $\mathcal{R}(D^*)$ with a semileptonic tagging method}},
  \href{https://doi.org/10.1103/PhysRevLett.124.161803}{\emph{Phys. Rev. Lett.}
  {\bfseries 124} (2020) 161803}
  [\href{https://arxiv.org/abs/1910.05864}{{\ttfamily 1910.05864}}].

\bibitem{LHCb:2023zxo}
{\scshape LHCb} collaboration, \emph{{Measurement of the ratios of branching
  fractions $\mathcal{R}(D^{*})$ and $\mathcal{R}(D^{0})$}},
  \href{https://arxiv.org/abs/2302.02886}{{\ttfamily 2302.02886}}.

\bibitem{Belle:2015qfa}
{\scshape Belle} collaboration, \emph{{Measurement of the branching ratio of
  $\bar{B} \to D^{(\ast)} \tau^- \bar{\nu}_\tau$ relative to $\bar{B} \to
  D^{(\ast)} \ell^- \bar{\nu}_\ell$ decays with hadronic tagging at Belle}},
  \href{https://doi.org/10.1103/PhysRevD.92.072014}{\emph{Phys. Rev. D}
  {\bfseries 92} (2015) 072014}
  [\href{https://arxiv.org/abs/1507.03233}{{\ttfamily 1507.03233}}].

\bibitem{Belle:2016ure}
{\scshape Belle} collaboration, \emph{{Measurement of the branching ratio of
  $\bar{B}^0 \rightarrow D^{*+} \tau^- \bar{\nu}_{\tau}$ relative to $\bar{B}^0
  \rightarrow D^{*+} \ell^- \bar{\nu}_{\ell}$ decays with a semileptonic
  tagging method}},
  \href{https://doi.org/10.1103/PhysRevD.94.072007}{\emph{Phys. Rev. D}
  {\bfseries 94} (2016) 072007}
  [\href{https://arxiv.org/abs/1607.07923}{{\ttfamily 1607.07923}}].

\bibitem{LHCbSem}
{\scshape LHCb} collaboration, ``\emph{Measurement of $R(D^*)$ with hadronic
  $\tau^+$ decays at $\sqrt{s}= 13$ TeV by the LHCb collaboration}.''
  \url{https://indico.cern.ch/event/1231797/}.

\bibitem{HFLAV:2022pwe}
{\scshape HFLAV} collaboration, \emph{{Averages of $b$-hadron, $c$-hadron, and
  $\tau$-lepton properties as of 2021}},
  \href{https://arxiv.org/abs/2206.07501}{{\ttfamily 2206.07501}}.

\bibitem{Bigi:2016mdz}
D.~Bigi and P.~Gambino, \emph{{Revisiting $B\to D \ell \nu$}},
  \href{https://doi.org/10.1103/PhysRevD.94.094008}{\emph{Phys. Rev. D}
  {\bfseries 94} (2016) 094008}
  [\href{https://arxiv.org/abs/1606.08030}{{\ttfamily 1606.08030}}].

\bibitem{Bernlochner:2017jka}
F.U.~Bernlochner, Z.~Ligeti, M.~Papucci and D.J.~Robinson, \emph{{Combined
  analysis of semileptonic $B$ decays to $D$ and $D^*$: $R(D^{(*)})$,
  $|V_{cb}|$, and new physics}},
  \href{https://doi.org/10.1103/PhysRevD.95.115008}{\emph{Phys. Rev. D}
  {\bfseries 95} (2017) 115008}
  [\href{https://arxiv.org/abs/1703.05330}{{\ttfamily 1703.05330}}].

\bibitem{Jaiswal:2017rve}
S.~Jaiswal, S.~Nandi and S.K.~Patra, \emph{{Extraction of $|V_{cb}|$ from $B\to
  D^{(*)}\ell\nu_\ell$ and the Standard Model predictions of $R(D^{(*)})$}},
  \href{https://doi.org/10.1007/JHEP12(2017)060}{\emph{JHEP} {\bfseries 12}
  (2017) 060} [\href{https://arxiv.org/abs/1707.09977}{{\ttfamily
  1707.09977}}].

\bibitem{Gambino:2019sif}
P.~Gambino, M.~Jung and S.~Schacht, \emph{{The $V_{cb}$ puzzle: An update}},
  \href{https://doi.org/10.1016/j.physletb.2019.06.039}{\emph{Phys. Lett. B}
  {\bfseries 795} (2019) 386}
  [\href{https://arxiv.org/abs/1905.08209}{{\ttfamily 1905.08209}}].

\bibitem{Bordone:2019vic}
M.~Bordone, M.~Jung and D.~van Dyk, \emph{{Theory determination of $\bar{B}\to
  D^{(*)}\ell^-\bar\nu$ form factors at $\mathcal{O}(1/m_c^2)$}},
  \href{https://doi.org/10.1140/epjc/s10052-020-7616-4}{\emph{Eur. Phys. J. C}
  {\bfseries 80} (2020) 74} [\href{https://arxiv.org/abs/1908.09398}{{\ttfamily
  1908.09398}}].

\bibitem{Martinelli:2021onb}
G.~Martinelli, S.~Simula and L.~Vittorio, \emph{{$\vert V_{cb} \vert$ and
  $R(D^{(*)})$ using lattice QCD and unitarity}},
  \href{https://doi.org/10.1103/PhysRevD.105.034503}{\emph{Phys. Rev. D}
  {\bfseries 105} (2022) 034503}
  [\href{https://arxiv.org/abs/2105.08674}{{\ttfamily 2105.08674}}].

\bibitem{Blanke:2018yud}
M.~Blanke, A.~Crivellin, S.~de~Boer, T.~Kitahara, M.~Moscati, U.~Nierste
  et~al., \emph{{Impact of polarization observables and $ B_c\to \tau \nu$ on
  new physics explanations of the $b\to c \tau \nu$ anomaly}},
  \href{https://doi.org/10.1103/PhysRevD.99.075006}{\emph{Phys. Rev. D}
  {\bfseries 99} (2019) 075006}
  [\href{https://arxiv.org/abs/1811.09603}{{\ttfamily 1811.09603}}].

\bibitem{Blanke:2019qrx}
M.~Blanke, A.~Crivellin, T.~Kitahara, M.~Moscati, U.~Nierste and
  I.~Ni\v{s}and\v{z}i\'c, \emph{{Addendum to \textquotedblleft{}Impact of
  polarization observables and $B_c\to \tau \nu$ on new physics explanations of
  the $b\to c \tau \nu$ anomaly''}},
  \href{https://doi.org/10.1103/PhysRevD.100.035035}{\emph{Phys. Rev. D}
  {\bfseries 100} (2019) 035035}
  [\href{https://arxiv.org/abs/1905.08253}{{\ttfamily 1905.08253}}].

\bibitem{LHCb:2022piu}
{\scshape LHCb} collaboration, \emph{{Observation of the decay $
  \Lambda_b^0\rightarrow \Lambda_c^+\tau^-\overline{\nu}_{\tau}$}},
  \href{https://doi.org/10.1103/PhysRevLett.128.191803}{\emph{Phys. Rev. Lett.}
  {\bfseries 128} (2022) 191803}
  [\href{https://arxiv.org/abs/2201.03497}{{\ttfamily 2201.03497}}].

\bibitem{Fedele:2022iib}
M.~Fedele, M.~Blanke, A.~Crivellin, S.~Iguro, T.~Kitahara, U.~Nierste et~al.,
  \emph{{Impact of
  \ensuremath{\Lambda_b}\textrightarrow{}\ensuremath{\Lambda_c}\ensuremath{\tau}\ensuremath{\nu}
  measurement on new physics in b\textrightarrow{}c\ensuremath{\ell\nu}
  transitions}}, \href{https://doi.org/10.1103/PhysRevD.107.055005}{\emph{Phys.
  Rev. D} {\bfseries 107} (2023) 055005}
  [\href{https://arxiv.org/abs/2211.14172}{{\ttfamily 2211.14172}}].

\bibitem{Sakaki:2013bfa}
Y.~Sakaki, M.~Tanaka, A.~Tayduganov and R.~Watanabe, \emph{{Testing leptoquark
  models in $\bar B \to D^{(*)} \tau \bar\nu$}},
  \href{https://doi.org/10.1103/PhysRevD.88.094012}{\emph{Phys. Rev. D}
  {\bfseries 88} (2013) 094012}
  [\href{https://arxiv.org/abs/1309.0301}{{\ttfamily 1309.0301}}].

\bibitem{Hiller:2014yaa}
G.~Hiller and M.~Schmaltz, \emph{{$R_K$ and future $b \to s \ell \ell$ physics
  beyond the standard model opportunities}},
  \href{https://doi.org/10.1103/PhysRevD.90.054014}{\emph{Phys. Rev.}
  {\bfseries D90} (2014) 054014}
  [\href{https://arxiv.org/abs/1408.1627}{{\ttfamily 1408.1627}}].

\bibitem{Dorsner:2016wpm}
I.~Dor\v{s}ner, S.~Fajfer, A.~Greljo, J.F.~Kamenik and N.~Ko\v{s}nik,
  \emph{{Physics of leptoquarks in precision experiments and at particle
  colliders}}, \href{https://doi.org/10.1016/j.physrep.2016.06.001}{\emph{Phys.
  Rept.} {\bfseries 641} (2016) 1}
  [\href{https://arxiv.org/abs/1603.04993}{{\ttfamily 1603.04993}}].

\bibitem{Dumont:2016xpj}
B.~Dumont, K.~Nishiwaki and R.~Watanabe, \emph{{LHC constraints and prospects
  for $S_1$ scalar leptoquark explaining the $\bar B \to D^{(*)} \tau \bar\nu$
  anomaly}}, \href{https://doi.org/10.1103/PhysRevD.94.034001}{\emph{Phys. Rev.
  D} {\bfseries 94} (2016) 034001}
  [\href{https://arxiv.org/abs/1603.05248}{{\ttfamily 1603.05248}}].

\bibitem{Li:2016vvp}
X.-Q.~Li, Y.-D.~Yang and X.~Zhang, \emph{{Revisiting the one leptoquark
  solution to the R(D$^{(*)}$) anomalies and its phenomenological
  implications}}, \href{https://doi.org/10.1007/JHEP08(2016)054}{\emph{JHEP}
  {\bfseries 08} (2016) 054}
  [\href{https://arxiv.org/abs/1605.09308}{{\ttfamily 1605.09308}}].

\bibitem{Hiller:2016kry}
G.~Hiller, D.~Loose and K.~Sch{\"o}nwald, \emph{{Leptoquark Flavor Patterns \&
  B Decay Anomalies}},
  \href{https://doi.org/10.1007/JHEP12(2016)027}{\emph{JHEP} {\bfseries 12}
  (2016) 027} [\href{https://arxiv.org/abs/1609.08895}{{\ttfamily
  1609.08895}}].

\bibitem{Bhattacharya:2016mcc}
B.~Bhattacharya, A.~Datta, J.-P.~Gu\'evin, D.~London and R.~Watanabe,
  \emph{{Simultaneous Explanation of the $R_K$ and $R_{D^{(*)}}$ Puzzles: a
  Model Analysis}}, \href{https://doi.org/10.1007/JHEP01(2017)015}{\emph{JHEP}
  {\bfseries 01} (2017) 015}
  [\href{https://arxiv.org/abs/1609.09078}{{\ttfamily 1609.09078}}].

\bibitem{Chen:2017hir}
C.-H.~Chen, T.~Nomura and H.~Okada, \emph{{Excesses of muon $g-2$,
  $R_{D^{(\ast)}}$, and $R_K$ in a leptoquark model}},
  \href{https://doi.org/10.1016/j.physletb.2017.10.005}{\emph{Phys. Lett. B}
  {\bfseries 774} (2017) 456}
  [\href{https://arxiv.org/abs/1703.03251}{{\ttfamily 1703.03251}}].

\bibitem{Crivellin:2017zlb}
A.~Crivellin, D.~M{\"u}ller and T.~Ota, \emph{{Simultaneous explanation of
  R(D$^{(*)}$) and $b\to s\mu^{+}\mu^{-}$: the last scalar leptoquarks
  standing}}, \href{https://doi.org/10.1007/JHEP09(2017)040}{\emph{JHEP}
  {\bfseries 09} (2017) 040}
  [\href{https://arxiv.org/abs/1703.09226}{{\ttfamily 1703.09226}}].

\bibitem{Jung:2018lfu}
M.~Jung and D.M.~Straub, \emph{{Constraining new physics in $b\to c\ell\nu$
  transitions}}, \href{https://doi.org/10.1007/JHEP01(2019)009}{\emph{JHEP}
  {\bfseries 01} (2019) 009}
  [\href{https://arxiv.org/abs/1801.01112}{{\ttfamily 1801.01112}}].

\bibitem{Aydemir:2019ynb}
U.~Aydemir, T.~Mandal and S.~Mitra, \emph{{Addressing the ${\mathbf
  R_{D^{(*)}}}$ anomalies with an ${\mathbf S_1}$ leptoquark from
  $\mathbf{SO(10)}$ grand unification}},
  \href{https://doi.org/10.1103/PhysRevD.101.015011}{\emph{Phys. Rev. D}
  {\bfseries 101} (2020) 015011}
  [\href{https://arxiv.org/abs/1902.08108}{{\ttfamily 1902.08108}}].

\bibitem{Popov:2019tyc}
O.~Popov, M.A.~Schmidt and G.~White, \emph{{$R_2$ as a single leptoquark
  solution to $R_{D^{(*)}}$ and $R_{K^{(*)}}$}},
  \href{https://doi.org/10.1103/PhysRevD.100.035028}{\emph{Phys. Rev. D}
  {\bfseries 100} (2019) 035028}
  [\href{https://arxiv.org/abs/1905.06339}{{\ttfamily 1905.06339}}].

\bibitem{Crivellin:2019dwb}
A.~Crivellin, D.~M\"uller and F.~Saturnino, \emph{{Flavor Phenomenology of the
  Leptoquark Singlet-Triplet Model}},
  \href{https://doi.org/10.1007/JHEP06(2020)020}{\emph{JHEP} {\bfseries 06}
  (2020) 020} [\href{https://arxiv.org/abs/1912.04224}{{\ttfamily
  1912.04224}}].

\bibitem{Iguro:2020keo}
S.~Iguro, M.~Takeuchi and R.~Watanabe, \emph{{Testing leptoquark/EFT in
  ${\bar{B}} \rightarrow {D^{(*)}}l{\bar{\nu }}$ at the LHC}},
  \href{https://doi.org/10.1140/epjc/s10052-021-09125-5}{\emph{Eur. Phys. J. C}
  {\bfseries 81} (2021) 406}
  [\href{https://arxiv.org/abs/2011.02486}{{\ttfamily 2011.02486}}].

\bibitem{Athron:2021iuf}
P.~Athron, C.~Bal\'azs, D.H.J.~Jacob, W.~Kotlarski, D.~St\"ockinger and
  H.~St\"ockinger-Kim, \emph{{New physics explanations of $a_{\mu}$ in light of
  the FNAL muon g \ensuremath{-} 2 measurement}},
  \href{https://doi.org/10.1007/JHEP09(2021)080}{\emph{JHEP} {\bfseries 09}
  (2021) 080} [\href{https://arxiv.org/abs/2104.03691}{{\ttfamily
  2104.03691}}].

\bibitem{Pendleton:1980as}
B.~Pendleton and G.G.~Ross, \emph{{Mass and Mixing Angle Predictions from
  Infrared Fixed Points}},
  \href{https://doi.org/10.1016/0370-2693(81)90017-4}{\emph{Phys. Lett. B}
  {\bfseries 98} (1981) 291}.

\bibitem{Ciuchini:2022wbq}
M.~Ciuchini, M.~Fedele, E.~Franco, A.~Paul, L.~Silvestrini and M.~Valli,
  \emph{{Constraints on lepton universality violation from rare B decays}},
  \href{https://doi.org/10.1103/PhysRevD.107.055036}{\emph{Phys. Rev. D}
  {\bfseries 107} (2023) 055036}
  [\href{https://arxiv.org/abs/2212.10516}{{\ttfamily 2212.10516}}].

\bibitem{Greljo:2022jac}
A.~Greljo, J.~Salko, A.~Smolkovi\v{c} and P.~Stangl, \emph{{Rare b decays meet
  high-mass Drell-Yan}},
  \href{https://doi.org/10.1007/JHEP05(2023)087}{\emph{JHEP} {\bfseries 05}
  (2023) 087} [\href{https://arxiv.org/abs/2212.10497}{{\ttfamily
  2212.10497}}].

\bibitem{Alguero:2023jeh}
M.~Alguer\'o, A.~Biswas, B.~Capdevila, S.~Descotes-Genon, J.~Matias and
  M.~Novoa-Brunet, \emph{{To (b)e or not to (b)e: No electrons at LHCb}},
  \href{https://arxiv.org/abs/2304.07330}{{\ttfamily 2304.07330}}.

\bibitem{Buras:2014fpa}
A.J.~Buras, J.~Girrbach-Noe, C.~Niehoff and D.M.~Straub, \emph{{$ B\to
  {K}^{\left(\ast \right)}\nu \overline{\nu} $ decays in the Standard Model and
  beyond}}, \href{https://doi.org/10.1007/JHEP02(2015)184}{\emph{JHEP}
  {\bfseries 02} (2015) 184} [\href{https://arxiv.org/abs/1409.4557}{{\ttfamily
  1409.4557}}].

\bibitem{Grygier:2017tzo}
{\scshape Belle} collaboration, \emph{{Search for $\boldsymbol{B\to
  h\nu\bar{\nu}}$ decays with semileptonic tagging at Belle}},
  \href{https://doi.org/10.1103/PhysRevD.97.099902,
  10.1103/PhysRevD.96.091101}{\emph{Phys. Rev.} {\bfseries D96} (2017) 091101}
  [\href{https://arxiv.org/abs/1702.03224}{{\ttfamily 1702.03224}}].

\bibitem{BelleIISemKnunu}
{\scshape Belle II} collaboration, ``Recent belle ii results on radiative and
  electroweak penguin decays.''
  \url{https://indico.desy.de/event/34916/contributions/146877}.

\bibitem{Belle-II:2018jsg}
{\scshape Belle-II} collaboration, \emph{{The Belle II Physics Book}},
  \href{https://doi.org/10.1093/ptep/ptz106}{\emph{PTEP} {\bfseries 2019}
  (2019) 123C01} [\href{https://arxiv.org/abs/1808.10567}{{\ttfamily
  1808.10567}}].

\bibitem{Buchmuller:1985jz}
W.~Buchmuller and D.~Wyler, \emph{Effective lagrangian analysis of new
  interactions and flavor conservation},
  \href{https://doi.org/10.1016/0550-3213(86)90262-2}{\emph{Nucl.Phys.B}
  {\bfseries 268} (1986) 621}.

\bibitem{Grzadkowski:2010es}
B.~Grzadkowski, M.~Iskrzynski, M.~Misiak and J.~Rosiek, \emph{Dimension-six
  terms in the standard model lagrangian},
  \href{https://doi.org/10.1007/JHEP10(2010)085}{\emph{JHEP} {\bfseries 10}
  (2010) 085} [\href{https://arxiv.org/abs/1008.4884}{{\ttfamily 1008.4884}}].

\bibitem{Alonso:2014csa}
R.~Alonso, B.~Grinstein and J.~Martin~Camalich, \emph{{$SU(2)\times U(1)$ gauge
  invariance and the shape of new physics in rare $B$ decays}},
  \href{https://doi.org/10.1103/PhysRevLett.113.241802}{\emph{Phys. Rev. Lett.}
  {\bfseries 113} (2014) 241802}
  [\href{https://arxiv.org/abs/1407.7044}{{\ttfamily 1407.7044}}].

\bibitem{Aebischer:2015fzz}
J.~Aebischer, A.~Crivellin, M.~Fael and C.~Greub, \emph{{Matching of gauge
  invariant dimension-six operators for $b\to s$ and $b\to c$ transitions}},
  \href{https://doi.org/10.1007/JHEP05(2016)037}{\emph{JHEP} {\bfseries 05}
  (2016) 037} [\href{https://arxiv.org/abs/1512.02830}{{\ttfamily
  1512.02830}}].

\bibitem{Iguro:2022yzr}
S.~Iguro, T.~Kitahara and R.~Watanabe, \emph{{Global fit to $b \to c\tau\nu$
  anomalies 2022 mid-autumn}},
  \href{https://arxiv.org/abs/2210.10751}{{\ttfamily 2210.10751}}.

\bibitem{Becirevic:2018afm}
D.~Be{\v c}irevi{\'c}, I.~Dor{\v s}ner, S.~Fajfer, N.~Ko{\v s}nik,
  D.A.~Faroughy and O.~Sumensari, \emph{{Scalar leptoquarks from grand unified
  theories to accommodate the $B$-physics anomalies}},
  \href{https://doi.org/10.1103/PhysRevD.98.055003}{\emph{Phys. Rev.}
  {\bfseries D98} (2018) 055003}
  [\href{https://arxiv.org/abs/1806.05689}{{\ttfamily 1806.05689}}].

\bibitem{Buchmuller:1986zs}
W.~Buchmuller, R.~Ruckl and D.~Wyler, \emph{{Leptoquarks in Lepton - Quark
  Collisions}}, \href{https://doi.org/10.1016/0370-2693(87)90637-X}{\emph{Phys.
  Lett. B} {\bfseries 191} (1987) 442}.

\bibitem{Angelescu:2018tyl}
A.~Angelescu, D.~Be{\v c}irevi{\'c}, D.A.~Faroughy and O.~Sumensari,
  \emph{{Closing the window on single leptoquark solutions to the $B$-physics
  anomalies}}, \href{https://doi.org/10.1007/JHEP10(2018)183}{\emph{JHEP}
  {\bfseries 10} (2018) 183}
  [\href{https://arxiv.org/abs/1808.08179}{{\ttfamily 1808.08179}}].

\bibitem{Gonzalez-Alonso:2017iyc}
M.~Gonz\'alez-Alonso, J.~Martin~Camalich and K.~Mimouni,
  \emph{{Renormalization-group evolution of new physics contributions to
  (semi)leptonic meson decays}},
  \href{https://doi.org/10.1016/j.physletb.2017.07.003}{\emph{Phys. Lett. B}
  {\bfseries 772} (2017) 777}
  [\href{https://arxiv.org/abs/1706.00410}{{\ttfamily 1706.00410}}].

\bibitem{Aebischer:2018acj}
J.~Aebischer, A.~Crivellin and C.~Greub, \emph{{QCD improved matching for
  semileptonic B decays with leptoquarks}},
  \href{https://doi.org/10.1103/PhysRevD.99.055002}{\emph{Phys. Rev. D}
  {\bfseries 99} (2019) 055002}
  [\href{https://arxiv.org/abs/1811.08907}{{\ttfamily 1811.08907}}].

\bibitem{Angelescu:2021lln}
A.~Angelescu, D.~Be\v{c}irevi\'c, D.A.~Faroughy, F.~Jaffredo and O.~Sumensari,
  \emph{{Single leptoquark solutions to the B-physics anomalies}},
  \href{https://doi.org/10.1103/PhysRevD.104.055017}{\emph{Phys. Rev. D}
  {\bfseries 104} (2021) 055017}
  [\href{https://arxiv.org/abs/2103.12504}{{\ttfamily 2103.12504}}].

\bibitem{Machacek:1983tz}
M.E.~Machacek and M.T.~Vaughn, \emph{{Two Loop Renormalization Group Equations
  in a General Quantum Field Theory. 1. Wave Function Renormalization}},
  \href{https://doi.org/10.1016/0550-3213(83)90610-7}{\emph{Nucl. Phys. B}
  {\bfseries 222} (1983) 83}.

\bibitem{Machacek:1983fi}
M.E.~Machacek and M.T.~Vaughn, \emph{{Two Loop Renormalization Group Equations
  in a General Quantum Field Theory. 2. Yukawa Couplings}},
  \href{https://doi.org/10.1016/0550-3213(84)90533-9}{\emph{Nucl. Phys. B}
  {\bfseries 236} (1984) 221}.

\bibitem{Banik:2023ogi}
S.~Banik and A.~Crivellin, \emph{{Renormalization Group Evolution with Scalar
  Leptoquarks}},  \href{https://arxiv.org/abs/2307.06800}{{\ttfamily
  2307.06800}}.

\bibitem{Bandyopadhyay:2021kue}
P.~Bandyopadhyay, S.~Jangid and A.~Karan, \emph{{Constraining scalar doublet
  and triplet leptoquarks with vacuum stability and perturbativity}},
  \href{https://doi.org/10.1140/epjc/s10052-022-10418-6}{\emph{Eur. Phys. J. C}
  {\bfseries 82} (2022) 516}
  [\href{https://arxiv.org/abs/2111.03872}{{\ttfamily 2111.03872}}].

\bibitem{Kowalska:2020gie}
K.~Kowalska, E.M.~Sessolo and Y.~Yamamoto, \emph{{Flavor anomalies from
  asymptotically safe gravity}},
  \href{https://doi.org/10.1140/epjc/s10052-021-09072-1}{\emph{Eur. Phys. J. C}
  {\bfseries 81} (2021) 272}
  [\href{https://arxiv.org/abs/2007.03567}{{\ttfamily 2007.03567}}].

\bibitem{Marzocca:2018wcf}
D.~Marzocca, \emph{{Addressing the B-physics anomalies in a fundamental
  Composite Higgs Model}},
  \href{https://doi.org/10.1007/JHEP07(2018)121}{\emph{JHEP} {\bfseries 07}
  (2018) 121} [\href{https://arxiv.org/abs/1803.10972}{{\ttfamily
  1803.10972}}].

\bibitem{FileviezPerez:2013zmv}
P.~Fileviez~Perez and M.B.~Wise, \emph{{Low Scale Quark-Lepton Unification}},
  \href{https://doi.org/10.1103/PhysRevD.88.057703}{\emph{Phys. Rev. D}
  {\bfseries 88} (2013) 057703}
  [\href{https://arxiv.org/abs/1307.6213}{{\ttfamily 1307.6213}}].

\bibitem{Varzielas:2015iva}
I.~de~Medeiros~Varzielas and G.~Hiller, \emph{{Clues for flavor from rare
  lepton and quark decays}},
  \href{https://doi.org/10.1007/JHEP06(2015)072}{\emph{JHEP} {\bfseries 06}
  (2015) 072} [\href{https://arxiv.org/abs/1503.01084}{{\ttfamily
  1503.01084}}].

\end{thebibliography}\endgroup
\end{document}